\documentclass[12pt,letterpaper]{article}
\usepackage{hyperref}
\usepackage{latexsym}
\usepackage{amssymb,amsfonts,amsmath}
\usepackage{graphicx}
\usepackage{indentfirst}
\usepackage{amsmath}
\usepackage{amsfonts}
\usepackage{amssymb}%
\usepackage{comment}
\usepackage{mathtools}
\usepackage{amsthm}
\usepackage{enumerate}
\usepackage{extarrows}

\usepackage[cmtip,all]{xy}
\usepackage{verbatim}
\usepackage{tikz-cd}
\usepackage{mathtools}
\usepackage{mathrsfs}
\usepackage{geometry}

\DeclareMathAlphabet{\mathpzc}{OT1}{pzc}{m}{it}

\newcommand{\beq}{\begin{equation}}
\newcommand{\eeq}{\end{equation}}
\newcommand{\bear}{\begin{eqnarray}}
\newcommand{\eear}{\end{eqnarray}}

\newcommand*{\defeq}{\mathrel{\vcenter{\baselineskip0.5ex \lineskiplimit0pt
                     \hbox{\scriptsize.}\hbox{\scriptsize.}}}%
                     =}

\newcommand{\slantone}[2]{{\raisebox{.1em}{$#1$}\left/\raisebox{-.1em}{$#2$}\right.}}

\ifx\pdfoutput\undefined
\else
\fi
\hypersetup{colorlinks=false,bookmarksopen,bookmarksnumbered,citecolor=blue,
pdfstartview=FitH}

\parskip=\medskipamount

\def\a{\alpha}

\def\b{\beta}

\def\g{\gamma}

\newcommand{\be}{\begin{equation}}
\newcommand{\ee}{\end{equation}}
\newcommand{\bea}{\begin{eqnarray}}
\newcommand{\eea}{\end{eqnarray}}

\newcommand{\ba}{\begin{array}}
\newcommand{\ea}{\end{array}}

\def\double #1{#1{\hbox{\kern-2pt $#1$}}}

\newcommand{\bsubeq}{\begin{subequations}}
\newcommand{\esubeq}{\end{subequations}}

\newcommand{\virgolette}{``}

\newcommand{\mani}{\ensuremath{\mathpzc{M}}}

\theoremstyle{definition}

\theoremstyle{plain}

\begin{document}

\begin{titlepage}
\begin{flushright}
\par\end{flushright}
\vskip 0.5cm
\begin{center}
\textbf{\LARGE \bf Cohomology of Lie Superalgebras: \\ Forms, Integral Forms and 
Coset Superspaces}

\vskip 1cm

\large {\bf R.~Catenacci}$^{~a,b,c,}$\footnote{roberto.catenacci@uniupo.it}
\large {\bf C.~A.~Cremonini}$^{~d,e,}$\footnote{carlo.alberto.cremonini@gmail.com}, \\
\large {\bf P.~A.~Grassi}$^{~b,c,f,}$\footnote{pietro.grassi@uniupo.it},
\large {\bf S.~Noja}$^{~g,}$\footnote{noja@mathi.uni-heidelberg.de}\,.  

\vskip .5cm {
\small
\centerline{$^{(a)}${\it Gruppo Nazionale di Fisica Matematica, InDAM, Piazzale Aldo Moro 5, 00185, Roma}}
\centerline{$^{(b)}$
\it Dipartimento di Scienze e Innovazione Tecnologica (DiSIT),} 
\centerline{\it Universit\`a del Piemonte Orientale, viale T.~Michel, 11, 15121 Alessandria, Italy}
\centerline{$^{(c)}$
\it Arnold-Regge Center, via P.~Giuria 1,  10125 Torino, Italy}
\centerline{$^{(d)}$ \it Dipartimento di Scienze e Alta Tecnologia (DiSAT),}
\centerline{\it Universit\`a degli Studi dell'Insubria, via Valleggio 11, 22100 Como, Italy}
\centerline{$^{(e)}$ \it INFN, Sezione di Milano, via G.~Celoria 16, 20133 Milano, Italy} 
\centerline{$^{(f)}$
\it INFN, Sezione di Torino, via P.~Giuria 1, 10125 Torino, Italy}
\centerline{$^{(g)}$ 
\it Mathematisches Institut, Universit\"{a}t Heidelberg, Germany}
}
\end{center}

\begin{abstract}
\noindent We study Chevalley-Eilenberg cohomology of physically relevant Lie superalgebras related to supersymmetric theories, providing explicit expressions for their cocycles in terms of their Maurer-Cartan forms. 
We then include integral forms in the picture by defining a notion of integral forms related to a Lie superalgebra. We develop a suitable generalization of Chevalley-Eilenberg cohomology extended to integral forms and we prove that it is isomorphic to the ordinary Chevalley-Eilenberg cohomology of the Lie superalgebra. Next we study equivariant Chevalley-Eilenberg cohomology for coset superspaces, which plays a crucial role in supergravity and superstring models. Again, we treat explicitly several examples, providing cocycles' expressions and revealing a characteristic infinite dimensional cohomology.
 
%We construct several interesting examples of Chevalley-Eilenberg (CE) cohomologies for 
%superalgebras and coset manifolds. Since the introduction of CE cohomologies in the 
%realm of supergravity and superstrings, there has been new developments in understanding 
%differential forms in supermanifolds. In particular, the integral forms became part of the common 
%language and therefore we study the CE cohomology in the sector of integral forms. We discover 
%an isomorphism between differential forms and integral forms cohomology and and infinite 
%dimensional cohomology classes for coset spaces. We clarify the construction of equivariant 
%CE cohomology in the case of superalgebras and we present several explicit examples. 
%\\
 \end{abstract}

\vfill{}
\vspace{1.5cm}
\end{titlepage}
\newpage\setcounter{footnote}{0}
\tableofcontents

\section{Introduction} \setcounter{equation}{0}

\noindent The mathematical development of cohomology of Lie algebras \cite{CE} \cite{HocSer} has been prompted and characterized by a twofold reason in relation to the theory of Lie groups.  \\
On one hand, in a diverging direction with respect to Lie groups, Lie algebra cohomology unties the representation theory of Lie algebras from the corresponding representation theory of Lie groups, by allowing a completely algebraic proof of the \emph{Weyl theorem} \cite{Weyl}, which was originally of analytic nature. On the other hand, in a converging direction with respect to Lie groups, in many important instances Lie algebra cohomology makes computations of the de Rham cohomology of the corresponding Lie groups easier. Nowadays, applications of Lie algebra cohomology range from representation theory in pure mathematics to modern physics - let us just recall that \emph{Kac-Moody} and \emph{Virasoro algebras}, which play central role in string theory, are central extensions of the \emph{polynomial loop-algebra} and the \emph{Witt algebra} respectively, and, as such, they are related to Lie algebra's 2-cohomology group. While it is quite natural to generalize a cohomology theory from Lie algebra to Lie superalgebra \cite{Kac1} \cite{Leites} (more recent reviews and computations can be found in \cite{deAzcarraga:1995jw,deAzcarraga:2001fi,deAzcarraga:2002xi,deAzcarraga:2004zj}).

both from a derived-functorial point of view and, more concretely, via cochain complexes, it can be seen that the two directions sketched above are meant to breakdown as one moves to the super setting. Indeed, in the representation theoretic direction, there is no Weyl theorem for Lie superalgebras, initially leading to the opinion that the cohomology theory is rather empty and meaningless. Further, in the topological direction, when working with Lie supergroups and their related Lie superalgebras, {Cartan theorem} resists to a naive \virgolette super'' generalization, as it only encodes topological informations. On the other hand a different point of view is possible, namely one can look at the the failure of Weyl theorem in the supersymmetric setting as an opportunity, rather than a pathological feature of the theory, for it suggests that the cohomology groups of Lie superalgebras might have a much richer structure than the one that can be guessed by analogy with the ordinary theory. Remarkably, physics is paving this way: cocycles arising from cohomology of Lie superalgebras - in particular, \emph{Poincar\'e superalgebras} - are getting related to higher \emph{Wess-Zumino-Witten} (\emph{WZW}) terms in supersymmetric Lagrangians (the so-called \emph{brane scan} and its recent higher-version, the \emph{brane bouquet}, which promotes Lie superalgebras to $L_{\infty}$-superalgebras and consider their cohomology) \cite{BaezHuer} \cite{Brandt} \cite{FSS} \cite{FSS2}. It is fair to observe, thought, that even the cohomology of a finite dimensional Lie superalgebra does not vanish in general for degree greater than the dimension of the algebra - as it happens in the ordinary case instead -: this makes the actual computation of the cohomology of Lie superalgebras into a very difficult task in general. Accordingly, results can be found in literature for specific choices of superalgebras - in particular in low-degree \cite{SZ} -, but only very few results encompassing the whole framework are available \cite{FL}, even just for the \emph{Betti numbers} of Lie superalgebras. %For example, to the best knowledge of the writer, there is \emph{no} reference reporting the \emph{Betti numbers} of the \emph{simple} (exceptional and non-exceptional) \emph{classical Lie superalgebras} \cite{Kac2}. 
Even less is known regarding the cohomology and the structure of cocycles of coset or homogeneous superspaces, which play a fundamental role in many superstring and supergravity models. If on one hand it is likely that a detailed knowledge of this equivariant cohomologies would help understanding the geometric nature and invariant structure of convoluted supergravity Langrangians \cite{Grozman:2002rk} \cite{Grozman}, it is also fair to notice that - once again - computations are difficult even in the most basic examples.

On a different note, getting back to the relations between algebras and groups, as mentioned above, it is a well-known fact that the de Rham cohomology of a Lie group can be formulated in terms of its underlying Lie algebra, thus making feasible computations otherwise very difficult. Whereas one tries to generalize this to Lie supergroups, (s)he would run into an issue, which is deeply ingrained in the theory of forms and the related integration theory in supergeometry. Indeed, in order to formulate a coherent notion of geometric integration on supermanifolds \cite{Leites:1980rna}, besides differential forms, one also need to take into account \emph{integral forms}, a notion which is crucial, thought not widely known and understood: for example, a supergeometric analogue of Stokes' theorem \cite{Manin} \cite{Witten} is proved using integral forms. On the other hand, it needs to be remarked that Lie superalgebra cohomology is nothing but a \virgolette $\mathbb{Z}_2$-graded generalization'' of the ordinary Lie algebra cohomology, and, as such, it is not capable to account for objects other than differential forms on supermanifolds, such as in particular, integral forms, which simply do not enter the picture \cite{YuZ}. It is natural to ask if it is possible to provide a formulation of Lie superalgebra cohomology capable of capturing properties of integral forms as well, and, in turn, what are the relations between the ordinary Lie superalgebra cohomology and this newly defined cohomology. 

In the present work, after a brief review of Chevalley-Eilenberg cohomology of Lie algebras and superalgebras and a basic introduction to integral forms - which aims at making the paper as self-consistent as possible -, we extend the notion of integral forms to a Lie superalgebraic context and we define a related notion of Chevalley-Eilenberg cohomology. We establish an isomorphism between the Chevalley-Eilenberg cohomology of integral forms of a superalgebra and the ordinary Chevalley-Eilenberg cohomology of the superalgebra in question. We then proceed to explicit computations of these cohomologies in several cases of physical interest, by looking at the Lie superalgebra of symmetries of relevant superspaces. However, it is fair to remark that, even if Lie supergroups - or supergroup manifolds, as they are called in the physics community - and their associated Lie superalgebras appear in several physical applications and have allowed to establish important results, coset supermanifolds actually open up to the most interesting and rich scenarios, offering several ways to take into account different amount of symmetries. For this reason, the last part of the paper is dedicated to the computations of equivariant Chevalley-Eilenberg cohomology for coset superspaces: several examples are discussed and typical phenomenology is pointed out.

\section{Chevalley-Eilenberg Cohomology: Main Definitions}

\subsection{Lie Algebras and Lie Superalgebras}

\noindent  We start providing the basic definitions, first in the usual setting, then in the \emph{super} one. Let $\mathfrak{g}$ be an ordinary finite dimensional Lie algebra defined over the field $k$, and let $V$ be a $\mathfrak{g}$-module or a representation space for $\mathfrak{g}.$ We define the (Chevalley-Eilenberg) $p$-cochains of $\mathfrak{g}$ valued in $V$ to be alternating $k$-linear maps from $\mathfrak{g}$ to $V$ \cite{CE},
\bear \label{cochains}
C_{CE}^p (\mathfrak{g}, V) \defeq Hom_k \left (\wedge^p  \mathfrak{g}, V \right ),
\eear
where we note in particular that $C_{CE}^0 (\mathfrak{g}, V) = Hom_k (k, V) \cong V$ for $p=0$ and where, in taking the exterior power $\mathfrak{g}$ is looked at as a vector space. Further, notice that if we take trivial coefficient, \emph{i.e.} $V = k$, as we will do in the rest of the paper, we simply have $C_{CE}^p (\mathfrak{g}, k) = \bigwedge^p \mathfrak{g}^\ast$. The above \eqref{cochains} can be lifted into a complex by introducing the (Chevalley-Eilenberg) differential $d_{\mathfrak{g}}^p : C_{CE}^p (\mathfrak{g}, V) \rightarrow C_{CE}^{p+1} (\mathfrak{g}, V),$ defined as
\begin{align} \label{differential}
d_{\mathfrak{g}}^p f (x_1 \wedge \ldots \wedge x_{p+1}) \defeq & \sum_{1 \leq i < j \leq p+1} (-1)^{i+j} f ([x_i, x_j] \wedge x_1\wedge \ldots \wedge \hat{x}_i \wedge \ldots \wedge \hat{x}_j \wedge \ldots \wedge x_{p+1}) + \nonumber \\
& + \sum_{i = 1}^{p+1} (-1)^{i+1} x_i \cdot f (x_1 \wedge \ldots \wedge \hat{x}_i \wedge \ldots \wedge x_{p+1}),
\end{align}
for $f \in Hom_k (\wedge^p \mathfrak{g}, V)$ and where the hatted entry is omitted. Once again notice that if $V$ is a trivial $\mathfrak{g}$-module, as in the case $V = k$, the second summand vanishes identically, so that one has
\begin{align}
d_{\mathfrak{g}}^p f (x_1 \wedge \ldots \wedge x_{p+1}) \defeq & \sum_{1 \leq i < j \leq p+1} (-1)^{i+j} f ([x_i, x_j] \wedge x_1\wedge \ldots \wedge \hat{x}_i \wedge \ldots \wedge \hat{x}_j \wedge \ldots \wedge x_{p+1})
\end{align}
It is not too hard to prove that $d^{p+1} \circ d^{p} = 0$, so that one can define the Chevalley-Eilenberg complex of $\mathfrak{g}$ valued in $V$ as the pair $(C_{CE}^\bullet (\mathfrak{g}, V), d^\bullet)$. Given this definition, the cohomology is defined in the usual way: we call \emph{Chevalley-Eilenberg cocycles} the elements of the vector space
\bear
Z_{CE}^p (\mathfrak{g}, V) \defeq \{ f \in C_{CE}^p (\mathfrak{g}, V): d^p f =0 \},
\eear 
and \emph{Chevalley-Eilenberg coboundaries} the elements in the vector space
\bear
B_{CE}^{p} (\mathfrak{g}, V) \defeq \{f \in C_{CE}^p (\mathfrak{g}, V) : \exists g \in C_{CE}^{p-1} (\mathfrak{g}, V)  : f = d^{p-1} g \},
\eear
and we define the \emph{Chevalley-Eilenberg $p$-cohomology group} of $\mathfrak{g}$ valued in $V$ as the quotient vector space
\bear
H_{CE}^p (\mathfrak{g}, V) \defeq \slantone{Z_{CE}^p (\mathfrak{g}, V)}{B_{CE}^p (\mathfrak{g}, V)}.
\eear
%This construction, via cochains, of a cohomology theory for a Lie algebra was first introduced by Chevalley and Eilenberg, and it has the perk of being very concrete and immediately accessible. It is not too hard, though, to see that it is indeed equivalent to the, more abstract, derived $Ext$-functor construction of the cohomology of Lie algebra, where $H^p (\mathfrak{g}, V) \defeq Ext^p_{{U} (\mathfrak{g})} (k, {V})$, with $U (\mathfrak{g})$ being the universal enveloping algebra of $\mathfrak{g}$. Briefly, to see this one has to actually computes the $Ext$-functor, for example using the fact that the Koszul complex $\mathcal{K}_\bullet (\mathfrak{g}) \defeq U (\mathfrak{g}) \otimes_k \wedge^\bullet \mathfrak{g}$ of the Lie algebra $\mathfrak{g}$ provides a (projective) resolution of $k$ as a $\mathfrak{g}$-module and then apply the functor $Hom_{U (\mathfrak{g})} ( - , V)$ to the resolution $\mathcal{K}_{\bullet} (\mathfrak{g}) \rightarrow k \rightarrow 0$. By adjuction,  $Hom_{U(\mathfrak{g})} (U(\mathfrak{g}) \otimes_k \wedge^p \mathfrak{g}, V) \cong Hom_k (\wedge^p \mathfrak{g}, V).$ \\
Denoting now $\mathfrak{g}$ a Lie \emph{super}algebra with $\mathfrak{g} = \mathfrak{g}_0 \oplus \mathfrak{g}_1$ its \emph{even} and \emph{odd} components in the $\mathbb{Z}_2$-grading, one can easily generalize the above construction just by taking care of the signs related to the $\mathbb{Z}_2$-grading (\emph{parity}). In particular, the definition of cochains and cohomology groups is unchanged and the previous differential in \eqref{differential} modifies to \cite{Leites}
\begin{align} \label{differential}
d^p f (x_1 \wedge \ldots \wedge x_{p+1}) \defeq & \sum_{1 \leq i < j \leq p+1} (-1)^{i+ j + \delta_{i, j} + \delta_{i-1, j}} f ([x_i, x_j] \wedge x_1\wedge \ldots \wedge \hat{x}_i \wedge \ldots \hat{x}_j \wedge \ldots \wedge x_{p+1}) + \nonumber \\
& + \sum_{i = 1}^{p+1} (-1)^{i+1 + \delta_{r-1}, r} x_i \cdot f (x_1 \wedge \ldots \wedge \hat{x}_i \wedge \ldots \wedge x_{p+1}),
\end{align}
where $\delta_{i, j} \defeq |x_i| (|f| + \sum_{k=0}^i |x_k|) $ for any $f \in \bigwedge^p \mathfrak{g}^\ast \otimes V$, $x_i \in \mathfrak{g}$, in order to take into account the parity, \emph{i.e.}\ the $\mathbb{Z}_2$-grading of the elements. Also, notice that as soon as the odd dimension of the Lie superalgebra is greater than zero, \emph{i.e.} if $\mathfrak{g}$ is a true Lie superalgebra and not just a Lie algebra, the Chevalley-Eilenberg cochain complex is not bounded from above, in pretty much the same fashion of the de Rham complex of a supermanifold, \emph{i.e.}
\bear
C^{\bullet}_{CE} (\mathfrak{g}, V) \defeq \bigoplus_{p \in \mathbb{Z}} C^p_{CE}(\mathfrak{g}, V) \quad \mbox{with} \quad C^{p}_{CE} (\mathfrak{g}, V) \neq 0 \quad \forall p \geq0.
\eear

\noindent As previously mentioned, Chevalley-Eilenberg cohomology has made its entrance years ago in physics, in particular in the context of supergravity and more specifically in the \virgolette FDA'' (Free Differential Algebra) approach to supergravity due to D'Auria and Fre \cite{cube}. Construction of (semi) Free Differential Algebras, or $n/\infty$-Lie (super)algebras were indeed given iteratively in terms of Chevalley-Eilenberg cocycles of a given Lie (super)algebra. In some sense, because of its supergravity origin, this approach is closer to \emph{Cartan geometry} than the previous one, which has more algebraic taste. 

One starts with a Lie group $G$ - or \emph{group manifold} in the supergravity literature - and a $G$-module $V$, \emph{i.e.}\ a $k$-vector space endowed with an action $\rho : G \times V \rightarrow V$ of $G$ on $V$, such that $\rho_g \defeq \rho (g, \cdot) \in Aut_k (V)$ for any $g \in G.$ Now, an $n$-form on the Lie group $G$ valued in $V$, \emph{i.e.}\ an element of the vector space $\Omega^n (G, V) \defeq \Omega^n (G) \otimes_k V$, is said to be $G$-\emph{equivariant} if $\ell^\ast_g \omega = \rho_g \omega$ for any $g \in G$ and where $\ell_g : G \rightarrow G$ is the left translation by $g$. We call $\Omega^n (G, V)^{eq}$ the space of equivariant $n$-form valued in the $G$-module $V$. It is clear that a $G$-equivariant form is determined by its value at the origin on $G$, and in particular it can be proved that $\Omega^p (G, V)^{eq} \cong C_{CE}^p (\mathfrak{g}, V)$. Further, $(d\omega^{eq})_e = d_{\mathfrak{g}}\omega$, where $d_{\mathfrak{g}}$ is the Chevalley-Eilenberg differential \eqref{differential} and $d$ is the de Rham differential. This shows that the Lie algebra cohomology can be described in terms of the de Rham cohomology of (equivariant) differential forms on the Lie group whose the Lie algebra is associated, \emph{i.e.} $H^p (\mathfrak{g}, V) \cong H^p (\Omega^{p} (G, V)^{eq}, d)$, thus making contact between two seemingly different cohomologies and making possible to compute Lie algebra cohomology via \emph{forms}, see for example \cite{Knapp}.  \\

The above remarks are completely general. In order to make contact with the notation employed and results in the following sections, we will now look at the description of the Chevalley-Eilenberg cohomology in terms of forms in some more details in the case we will be concerned with, that of the trivial $\mathfrak{g}$-module $V=k$, where $k$ is the ground field. In this case we will simply write $C^p_{CE} (\mathfrak{g}) \defeq C^p_{CE} (\mathfrak{g}, k)$ for the Chevalley-Eilenberg cochains defined above, and we recall that $C^p_{CE} (\mathfrak{g}) = \bigwedge^p \mathfrak{g}^\ast$, see the definition \eqref{cochains}. Likewise, equivariance of forms becomes simply left-invariance, \emph{i.e.} the requirement $\ell^\ast_g \omega = \omega$. This means that denoting $\Omega^p_L (G)$ the vector space of left-invariant forms one has that the above isomorphism becomes $C^p_{CE} (\mathfrak{g}) \cong \Omega^p_L (G)$. Let us consider a $k$-basis of left-invariant forms $\omega^i \in \Omega^1_{L} (G)$ together with its dual basis of left-invariant vector field $X_i \in (\Omega^1_L(G))^\ast$, with $\omega^i_g (X_{j, g}) = \delta^{i}_j$ for any $g \in G$. Then, the $\omega^i \in \Omega^1_L (G)$ satisfy the \emph{Maurer-Cartan structure equation} 
\bear \label{MCeq1}
d \omega^i = -\frac{1}{2} C_{~jk}^i \, \omega^j \wedge \omega^k,
\eear   
where the $C_{~jk}^i $ are the \emph{structure constants} relative to the basis $\omega^i$. 
The sums over repeated indices are understood. These equations are equivalent to the Lie braket relations for the basis $X^i$ of the algebra of left-invariant vector fields, $[X_j, X_k ] = C_{jk}^i X_i$. Also it can be easily checked that $d\circ d = 0$ is equivalent to \emph{Jacobi identity}, as 
\begin{align}
d(d\omega^k) & = - \frac{1}{2}  C^{k}_{~ij} \, d\omega^i \wedge \omega^j +  \frac{1}{2} C^{k}_{~ij} \, \omega^i \wedge d \omega^j = \frac{1}{2} C^{k}_{~i[j} C^{i}_{~l m]}\, \omega^{l} \wedge \omega^m \wedge \omega^j = 0,
\end{align}
where $\omega^i \in \Omega_L (G)$ and where $C^{k}_{~i[j} C^{i}_{~l m]} = 0$ is indeed the Jacobi identity. These will be the fundamental ingredients to actually compute cohomologies (notice that the differential is a derivation, so that it extends to higher forms).\\ 

\noindent In the present paper we will deal only with \emph{matrix Lie groups}, \emph{i.e.} Lie groups which admit an embedding into some $GL$-group: in this case, the above is equivalent to take a basis of forms $\mathpzc{V} = dg g^{-1}$, where $g \defeq (g_{ij})$ is matrix-valued, which we call Maurer-Cartan forms, as they satisfy Maurer-Cartan equations \eqref{MCeq1} by construction. In turns, we will take the cochains to be generated starting from the basis of Maurer-Cartan forms $\{\mathpzc{V}^i\}$, \emph{i.e.} the \emph{vielbeins} in the physics literature, so that 
\bear
C^p (\mathfrak{g}) = \Omega^{p}_L (G) = \bigg \{ %\sum_{i_1, \ldots, i_p}
 c_{i_1, \ldots, i_p} \mathpzc{V}^{i_1} \wedge \ldots \wedge \mathpzc{V}^{i_p} \bigg \} \quad \mbox{for} \quad c_{i_1, \ldots, i_p}  \in k.
\eear
Notice that the above discussion is readily generalizable to the $\mathbb{Z}_2$-graded super-setting of a Lie \emph{super}group $\mathpzc{G}$ and its Lie \emph{super}algebra $\mathfrak{g}$, but a remark about the \emph{parity} is in order: indeed, instead of considering the vector bundle of forms, we will consider its parity reversed version $\Omega^1(\mathpzc{G}) \defeq \Pi \mathcal{T}^\ast(\mathpzc{G})$, as it is customary in supergeometry: notice that in this convention the de Rham differential $d$ is an \emph{odd} derivation. This leads to consider \emph{even} and \emph{odd} vielbeins $\{\psi^\alpha | \mathpzc{V}^i \}$ generating the $\mathbb{Z}_2$-graded vector space $\Omega^1_L (\mathpzc{G})$, where the even $\psi^\alpha$'s arise from odd coordinates and the odd $\mathpzc{V}^i$'s arise from even coordinates. What it is crucial to observe is that, accordingly, this should be related to the \emph{parity changed dual} of the Lie superalgebra $\Pi \mathfrak{g}^\ast$, that is at the level of the cochains one has \bear
C^\bullet (\Pi \mathfrak{g})\defeq S^\bullet \Pi \mathfrak{g}^\ast \cong \Omega^\bullet_L (\mathpzc{G}),
\eear where $S^\bullet$ is the supersymmetric product functor (symmetrization operation)  \cite{Manin}. Likewise, at the level of the differentials, the de Rham differential is extended to both even and odd coordinates. The commutators characterizing the algebra or, dually, the Maurer-Cartan equations, become \emph{supercommutators}. In particular, on the parity reversed algebra $\Pi \mathfrak{g}$, if $\pi X$ and $\pi Y \in \Pi \mathfrak{g}$ we put $[ \pi X , \pi Y \} \defeq [X, Y\} $ for $X$ and $Y$ in $\mathfrak{g}.$

\subsection{Integral Forms and Chevalley-Eilenberg Cohomology}

\subsubsection{A Primer of Integral Forms on Supermanifolds}

\noindent Given a supermanifold $\mani$, say of dimension $n|m$, differential forms in $\Omega^\bullet (\mani) $ are not enough to define a coherent notion of integration on $\mani$. This leads to the introduction of \emph{integral forms}, which are geometrically as important as differential forms, see \cite{Manin} and the recent papers \cite{CGM, Castellani:2014goa, Castellani:2015paa, GrassiMaccaferri, Noja, Catenacci:2018jjj, CatenacciGrassiNoja2, CGN, CA1, CA2, CGP1, CGP2, CCGN, CNR}. Loosely speaking, whereas differential forms lead to a consistent geometric integration on ordinary bosonic submanifolds (\emph{i.e.}\ sub-manifolds of codimension $p|m$) in $\mani$, integral forms plays the same role on sub-supermanifolds of codimension $p|0$ in $\mani$, and in particular, they control integration on $\mani$ itself. Notice that, even if it is often left understood or not stated, integral forms are ubiquitous in theoretical high energy physics: for example, the Lagrangian density of a supersymmetric theory in superspace is indeed a \emph{top} integral form. 
There are (at least) two ways to introduce integral forms, which we now briefly recall. \\

The first approach is to define integral forms as \emph{generalized functions} on $\mbox{Tot}\, \Pi \mathcal{T} (\mani)$ \cite{Witten}, that is elements $
\omega (x^1, \ldots, x^n, d\theta^1, \ldots, d\theta^m | \theta^1, \ldots, \theta^m, dx^1, \ldots dx^n) \in \Pi \mathcal{T} (\mani)
$, where $x^i|\theta^\alpha$ are local coordinate for $\mani$, which only allows a \emph{distributional dependence} supported in $d\theta^1 = \ldots = d\theta^m = 0$. Algebraically, integral forms can be (roughly) described as $\Omega^\bullet (\mani)$-modules generated over the set (of \emph{Dirac delta} distributions and their derivatives) $\{ \delta^{(r_1)} (d\theta^1)\wedge \ldots \wedge \delta^{(r_m)} (d\theta^m) \}$, for $r_i \geq 0$, together with the defining relations
\bear\label{intbypartsdelta}
d\theta^{\alpha}\delta^{(k)}(d\theta^{\alpha})=-k\delta^{(k-1)}(d\theta_{\alpha}) \quad \mbox{for} \quad k\geq0
\eear 
for any $\alpha = 1, \ldots, m$ and any $k \geq 0$, which are deduced analytically by integration by parts. Notice that the case $k=0$ tells that the expressions $d\theta^\alpha \delta^{(0)} (d\theta^\alpha) $ vanishes, so that the presence of the delta's can be seen as a localization in the locus $d\theta^\alpha = 0$ in $\mbox{Tot}\, \Pi \mathcal{T} (\mani)$. Locally, an integral form $\omega_{int}$ is written as a (generalized) tensor
\begin{align} \label{intf}
\omega_{int} (x, & d\theta | \theta, dx) = \nonumber \\ 
& =  \sum_{i=1}^n \sum_{j =1}^m\sum_{a_i \in \{ 0,1\}, r_j \geq 0} \omega
_{\lbrack a_{1}\dots a_{m}r_{1}\dots r
_{m}]} (x |\theta) (dx^1)^{a_{1}}\dots (dx^n)^{a_{m}}\delta^{(r_{1})}(d\theta^1 )\dots
\delta^{(r_{m})}(d\theta^m),
\end{align}
where all indices are antisymmetric (recalling that two delta's anticommute with each other), and where we note that there cannot be $d\theta$'s thanks to the above relations \eqref{intbypartsdelta}. In what follows we will say that an integral forms has \emph{picture} $m$, to mean that we are considering expressions that admits only a distributional dependence on \emph{all} of the $m$ coordinates $d\theta^1, \ldots, d\theta^m$ on $\mbox{Tot}\, \Pi \mathcal{T} (\mani)$. Further, with reference to the previous expression \ref{intf}, we assign a \emph{degree} to an integral form according to the definition  
\bear
\deg (\omega_{int}) \defeq \sum_{i = 1}^n a_j - \sum_{j=1}^m r_j,
\eear
so that we will say that an integral form has picture $m$ and degree $p \leq n.$ In particular, a \emph{top} integral form is an integral form of degree $n$, 
\bear \label{berdel}
\omega^{top}_{int} = \omega (x | \theta ) dx^1 \ldots dx^n \delta (d\theta^1) \ldots \delta (d\theta^m),
\eear 
and it can be checked that this expression has the transformation properties of a section of the Berezinian line bundle $\mathcal{B}er (\mani) \defeq \mathcal{B}er^\ast (\Pi \mathcal{T}^\ast (\mani))$ of the supermanifold $\mani$. Notice that all of the integral forms as in \refeq{intf} can be generated from the above \ref{berdel} by repeatedly acting with \emph{contractions} along (coordinate) vector fields, \emph{i.e.}
\bear \label{mgit}
\omega_{int}^{n-\ell} = \iota_{X_1}\ldots \iota_{X_\ell} \, \omega^{top}_{int}, 
\eear
where we recall that in particular, for the coordinate vector fields $\partial_{x^i} | \partial_{\theta^\alpha} $ one has that $|\iota_{\partial_{x^i}}| = 1$ and $|\iota_{\partial_{\theta^\alpha}}| = 0$. The modules of integrals forms are then structured into a complex letting $d$ operate as the usual de Rham differential on $\Omega^\bullet (\mani)$ and declaring that its action on the delta's, is trivial \emph{i.e.} posing $d (\delta (d\theta^\alpha) ) = 0$ for any $\alpha$.

In the second approach one defines integral forms of degree $p$ as sections of the vector bundle on $\mani$
\bear \label{intform2}
\Sigma^p (\mani) \defeq \mathcal{B}er (\mani) \otimes_{\mathcal{O}_\mani} (\Omega^{n-p} (\mani))^\ast = \mathcal{B}er (\mani) \otimes_{\mathcal{O}_\mani} S^{n-p} (\Pi \mathcal{T}(\mani)).
\eear
where $\mathcal{B}er (\mani)$ is the Berezinian line bundle of $\mani$ and $\Pi \mathcal{T} (\mani)$ the parity-reversed tangent bundle. The correspondence between integral forms in the different representations reads
\bear
\omega^{(n-\ell)} = \mathpzc{D} \otimes \left( \pi X_1 {\odot} \ldots \odot \pi X_\ell \right) \ \leftrightsquigarrow \ \omega^{(n-\ell)} = \iota_{X^1} \ldots \iota_{X^\ell} \omega^{top}_{int}
\eear
where $\mathpzc{D}$ is a section of $\mathcal{B}er(\mani)$ and $\pi X_1 {\odot} \ldots \odot \pi X_\ell $ is a section of $S^\ell \Pi \mathcal{T} (\mani)$, together with the correspondence of sections of Berezinian line bundle, or integral top forms, mentioned above, \emph{i.e.} $\omega^{top}_{int} \leftrightsquigarrow \mathpzc{D}$.
Clearly, given the above tensor product structure, defining a nilpotent differential acting as $\delta^p : \Sigma^p (\mani) \rightarrow \Sigma^{p+1} (\mani)$ is not at all trivial matter, as originally discussed in 
\cite{Manin} and recently realized in \cite{CNR}, but this can be done as getting a complex which will in general be  \emph{unbounded from below}
\begin{align}
\xymatrix{
\ldots  \ar[r] & \mathcal{B}er (\mani) \otimes S^{n-p} (\Pi \mathcal{T}(\mani))  \ar[r] & \ldots \ar[r] & \mathcal{B}er (\mani) \otimes  \Pi \mathcal{T}(\mani) \ar[r] & \mathcal{B}er (\mani) \ar[r] & 0.
}
\end{align}
Remarkably, these different approaches, which agree in terms of general results, complement each others. If on one hand this second approach is probably more suitable when it comes to deal with mathematical and foundational issues where well-definiteness is crucial, on the other hand the first approach proves quite more effective when it comes to actual computations, and for this reason is favoured in applications to theoretical physics. The different nature of these two approaches is mirrored, for example, in the proof of which is probably the most important result in the theory, \emph{i.e.}\ the (natural) isomorphism between the cohomology of differential form $H^{p}_{d} (\Omega^{\bullet} (\mani))$ and integral forms $H^p_{\delta} (\Sigma^\bullet (\mani) )$ on supermanifolds, namely introducing in the first approach the crucial notion of \emph{Picture Changing Operators} (see, e.g., \cite{CGN}), which maps differential to integral forms and vice-versa, and via a spectral sequence argument in the second approach \cite{CNR}.  

\subsubsection{Defining Chevalley-Eilenberg Cohomology of Integral Forms}

\noindent In this section we investigate to what extent, in the case the supermanifold $\mani$ is a Lie supergroup $\mathpzc{G}$ with Lie superalgebra $\mathfrak{g}$, it is possible to define a notion of \virgolette integral form'' and in particular a \virgolette Chevalley-Eilenberg cohomology'' of integral forms related to $\mathfrak{g}$. Notice that, as explained above, the Chevalley-Eilenberg cohomology can be analogously introduced as the cohomology of the vector (super)space of the left-invariant differential forms for a certain Lie group: since over a supermanifold differential forms need to be supplemented by integral forms, it can be expected that there must exist an analogous notion of cohomology of \emph{left-invariant integral}, better than differential, \emph{forms}.\\

The first of the two approaches presented above is probably more straightforward in this respect. One takes a basis of Maurer-Cartan forms $\{ \psi^\alpha | \mathpzc{V}^i \}$, or supervielbein, with even $\psi$'s and odd $\mathpzc{V}$'s and restrict to consider only integral forms written in terms of them. More precisely, if $\mathpzc{Y}_{i | \alpha} \defeq \{ \mathcal{P}_i |\mathcal{Q}_\alpha \}$ is the basis of generators of the Lie superalgebra $\mathfrak{g}$ which is dual (up to a parity shift) to the basis of the Maurer-Cartan forms above, so that $ \psi^\alpha (\pi \mathcal{Q}_\beta) = \delta^{\alpha}_\beta $ and $\mathpzc{V}^i (\pi \mathcal{P}_j) = \delta^i_j$, then the most general integral form on $\mathfrak{g}$ of degree $n-\ell$, see \eqref{mgit}, will be written as 
\bear
\omega^{n-\ell}_{\mathfrak{g}} = %\sum_{i_1, \ldots, i_\ell}
 \omega^{i_{1} \ldots i_{\ell} } \iota_{\mathpzc{Y}^{i_1}} \ldots \iota_{\mathpzc{Y}^{i_\ell}} \omega^{top}_{\mathfrak{g}},
\eear
for $\mathpzc{Y}$ spanning both even and odd dimensions of $\mathfrak{g}$ and the indices of the tensor $\omega^{i_{1} \ldots i_{\ell} }$ symmetrized or anti-symmetrized according to the parity of the related contraction (the sum over repeated indices is understood). In the above expression one fixes the integral top form up to a multiplicative constant to be
\bear\label{BerForm}
\omega^{top}_{\mathfrak{g}} = \mathpzc{V}^1\ldots \mathpzc{V}^n \delta (\psi^1) \ldots \delta ( \psi^m ), 
\eear
that is $\omega^{top}_\mathfrak{g}$ is again expressed only in terms of of the Maurer-Cartan forms, which makes it formally left-invariant. Having set this stage and in the light of the discussion in the previous subsection, one can therefore generalize the Maurer-Cartan differential as to act on integral forms in the following way
\bear \label{dintf1}
\omega^{n-\ell}_{\mathfrak{g}} \longmapsto d (\omega^{n-\ell}_{\mathfrak{g}}) \defeq \frac{1}{2} %\sum_{A, B, C}
C^{A}_{\; \; BC} (\pi \mathpzc{Y}^\ast)^B (\pi \mathpzc{Y}^\ast)^{C} \iota_{\mathpzc{Y}^A}  \bigg ( %\sum_{i_1, \ldots, i_\ell}
 \omega^{i_{1} \ldots i_{\ell} } \iota_{\mathpzc{Y}^{i_1}} \ldots \iota_{\mathpzc{Y}^{i_\ell}} \omega^{top}_{\mathfrak{g}} \bigg ),
\eear
where $C^{A}_{\; \; BC}$ are the structure constants of the Lie superalgebra $\mathfrak{g}$ and $A,B$ and $C$ 
are the cumulative indices for $i|\alpha$. 
 Notice that the right-hand ride of the \eqref{dintf1} defines indeed an integral form of degree $n-\ell + 1$ and, once again, that the differential is indeed nilpotent thanks to Jacobi identity for the Lie superalgebra $\mathfrak{g}.$ \\

Defining integral forms for $\mathfrak{g}$ in the second approach requires some further explanations: the discussion is somewhat formal, therefore the reader can skip to the next section at the first reading. One can proceed specializing the definition \eqref{intform2} to a Lie supergroup $\mathpzc{G}$, but first the question of how to intrinsically define a \emph{left-invariant Berezinian} needs to be addressed. One might start with an analogy with the ordinary case, where the \emph{Haar determinant} - which, integrated, gives the volume of a compact Lie group -, is constructed by taking the top exterior power of the left-invariant $1$-forms $\mbox{Span}_{\mathbb{R}} \{ \omega^1, \ldots, \omega^n\} = \Omega^1_L (G)$ over the $n$-dimensional ordinary Lie group $G$, \emph{i.e.}\ $\det (G) = \mathbb{R} \cdot \omega^1 \wedge \ldots \wedge \omega^n.$ This construction cannot be generalized in a straighforward manner, mainly because the Berezinian of a vector space is not a top-exterior form. On the other hand, there exists a less known construction of the Berezinian of a vector superspace via the cohomology of a suitable generalization of the Koszul complex (see the quite recent papers \cite{NR} and \cite{CNR}): this should not surprise, as also the determinant appears in the same way from the Koszul complex. More precisely, given a vector $\mathbb{R}$-superspace $V$ of dimension $n|m$, one finds that the cohomology of the (dual of the) Koszul complex is concentrated in degree $n$, \emph{i.e.}\ $Ext^n_{S^\bullet V^\ast} (\mathbb{R}, S^\bullet V^\ast) \cong \Pi^{n+m} \mathbb{R}$ and an automorphism $\phi \in Aut (V)$ induces an automorphism on $Ext^n_{S^\bullet V^\ast} (\mathbb{R}, S^\bullet V^\ast)$ which is just the multiplication by the Berezinian of the automorphism $\mbox{Ber} (\phi)$, so that one rightfully defines $\mbox{Ber} (V ) \defeq Ext^n_{S^\bullet V^\ast} (\mathbb{R}, S^\bullet V^\ast)$ \cite{NR}. The computation of this cohomology is particularly useful also because it gives the generator of the Berezinian of $V$ in terms of the generators of the vector space $V$. %{\bf More explanations? Less explanations? } 
Shifting from algebra to geometry, one defines the Berezinian of the supermanifold to be the Berezinian of the tangent bundle $\mathcal{T}(\mani)$, or analogously the dual of the Berezinian of the parity-reversed cotangent bundle $\Omega^1 (\mani)$ as above. One finds that the Berezinian line bundle is (locally) generated by the class 
\bear
\mathcal{B}er (\mani) \cong \mathcal{O}_\mani \cdot   [dx^1 \wedge \ldots \wedge dx^n \otimes \partial_{\theta^1} \wedge \ldots \wedge \partial_{\theta^m}]
\eear in the corresponding $\mathcal{E}xt$-sheaf, where $x^i | \theta^\alpha$ for $i=1,\ldots,n$ and $\alpha = 1, \ldots, m$ are local coordinates for the supermanifold $\mani$. \\
This is what is needed in order to write a corresponding \emph{left-invariant Berezinian}, or \emph{Haar Berezinian} for a Lie supergroup: it is enough to consider the left-invariant \emph{odd} vector fields, call them $\{ \Psi^{(\ell)}_1, \ldots, \Psi^{(\ell)}_m \}$, generating $\mathfrak{g}_1$ and the left-invariant \emph{odd} 1-forms, call them $\{ \omega^{(\ell)1}, \ldots, \omega^{(\ell)n} \}$, generating $\Pi \mathfrak{g}^\ast_1$: then the Haar Berezinian is generated over $\mathbb{R}$ by the expression 
\bear
\mbox{Ber}^{\mathpzc{H}} (\mathfrak{g}) \cong \mathbb{R} \cdot [\omega^{(\ell)1} \wedge \ldots \wedge \omega^{(\ell)n} \otimes \Psi^{(\ell)}_1 \wedge \ldots \wedge \Psi^{(\ell)}_m ].
\eear
Notice that this corresponds to the vector superspace of densities of the vector space underlying the Lie superalgebra $\mathfrak{g}.$  \virgolette Dually'' to ordinary Chevally-Eilenberg cochains for a Lie superalgebra, integral forms cochains can then be introduced into this Lie-algebraic framework by looking at the definition \eqref{intform2} as
\bear\label{IntFormPoly}
C_{CE, \mathpzc{int}}^p (\mathfrak{g}) \defeq \mbox{Ber}^{\mathpzc{H}} (\mathfrak{g}) \otimes S^{n-p} \Pi \mathfrak{g}, 
\eear
where we are exploiting the usual isomorphism between left invariant vector fields on a Lie supergroup $\mathpzc{G}$ and elements of its Lie algebra $\mathfrak{g}$. In order to distinguish between them we henceforth call \emph{differential} Chevalley-Eilenberg $p$-cochains the elements in the vector superspace $C^p_{CE, \mathpzc{dif}} (\mathfrak{g}) \defeq S^p \Pi \mathfrak{g}^\ast$ and \emph{integral} Chevalley-Eilenberg $p$-cochains the elements in the vector superspace $C^p_{CE, \mathpzc{int}} (\mathfrak{g}) \defeq \mbox{Ber}^{\mathpzc{H}} (\mathfrak{g}) \otimes S^{n-p} \Pi \mathfrak{g},$ as above.\\
In order to structure this into a true cochain complex, one has to introduce a nilpotent differential acting as $\delta^p : C^p_{CE, \mathpzc{int}} (\mathfrak{g}) \rightarrow C^{p+1}_{CE, \mathpzc{int}} (\mathfrak{g})$. We first extend the notion of Lie derivative, or \emph{supercommutator}, to the whole supersymmetric product $S^n \Pi \mathfrak{g}$, this can be done recursively as follows. 
Given $\mathpzc{X} \in \mathfrak{g}$, having already defined $\mathcal{L}_{\mathpzc{X}} : S^h \Pi \mathfrak{g} \rightarrow S^h \Pi \mathfrak{g}$ for $h<p$ we uniquely define the action of $\mathcal{L}_\mathpzc{X} $ on $S^p \Pi \mathfrak{g}$ via the relation
\bear \label{recursive}
\mathcal{L}_{\mathpzc{X}} (\langle \omega, \tau \rangle ) = \langle \mathcal{L}_\mathpzc{X} (\omega) , \tau \rangle + (-1)^{|\omega| |\mathpzc{X}|} \langle \omega , \mathcal{L}_\mathpzc{X} (\tau) \rangle 
\eear
for any $\omega \in S^{i>0}\Pi \mathfrak{g}^\ast$ and $\tau \in S^p \Pi \mathfrak{g}$, and where $\langle \cdot , \cdot \rangle$ is the duality pairing between $\Pi \mathfrak{g}^\ast$ and $\Pi \mathfrak{g}$, extended to higher tensor powers. Notice that from \eqref{recursive} it follows that %{\bf Calcoletto da mettere?}
\bear\label{LIEA}
\mathcal{L}_X (Y)= \pi [X , \pi Y] 
\eear
for any $Y \in \Pi \mathfrak{g}$, \emph{i.e.}\ the Lie derivative of a parity-reversed field is a commutator, as it should. We now use this to introduce a differential, namely we define the following odd operator
\bear\label{DiffOpeDelta}
\xymatrix@R=1.5pt{
\delta^p : C^p_{CE, \mathpzc{int}} (\mathfrak{g}) \ar[r] & C^{p+1}_{CE, \mathpzc{int}} \\
\mathpzc{D} \otimes \tau \ar@{|->}[r] & \delta^p (\mathpzc{D} \otimes \tau) = \mathpzc{D} \otimes \sum_{A} \iota_{\pi \mathpzc{X}^\ast_A} \mathcal{L}_{\mathpzc{X}_A} (\tau) 
}
\eear
where the index $A$ runs over both even and odd coordinates and where $\mathpzc{D}$ is a Haar Berezinian tensor density in $\mbox{Ber}^\mathpzc{H} (\mathfrak{g})$ and $\{ \mathpzc{X}_A \}$ are left-invariant vector fields generating $\mathfrak{g}$, so that hence $\{ \pi \mathpzc{X}^\ast_A \}$ are generators for $\Pi \mathfrak{g}^\ast$. Here $\iota_{\pi \mathpzc{X}^\ast_A}$ is the contraction with the form $\pi \mathpzc{X}^\ast_A$, so that the above can be re-written as
\bear \label{diff2}
\delta^p (\mathpzc{D} \otimes \tau) = \mathpzc{D} \otimes \sum_{A} \langle \pi \mathpzc{X}_A^\ast , \mathcal{L}_{\mathpzc{X}_A} (\tau) \rangle.
\eear
Nilpotency can be checked formally as
\begin{align}
\frac{1}{2}\{\delta, \delta \} &  = \sum_{A,B} (\iota_{\pi \mathpzc{X}^\ast_A} \mathcal{L}_{\mathpzc{X}_A} \iota_{\pi \mathpzc{X}^\ast_B} \mathcal{L}_{\mathpzc{X}_B} + \iota_{\pi \mathpzc{X}^\ast_B} \mathcal{L}_{\mathpzc{X}_B}\iota_{\pi \mathpzc{X}^\ast_A} \mathcal{L}_{\mathpzc{X}_A}) \nonumber  \\
& = \sum_{A,B}\left ( (-1)^{|\mathpzc{X}_A||\mathpzc{X}_B| + |\mathpzc{X}_A|} + (-1)^{|\mathpzc{X}_A||\mathpzc{X}_B| + |\mathpzc{X}_A| +1}\right ) \iota_{\pi \mathpzc{X}^\ast_A} \iota_{\pi \mathpzc{X}^\ast_B} \mathcal{L}_{\mathpzc{X}_A} \mathcal{L}_{\mathpzc{X}_B}=0.
\end{align}
We thus introduce the cochain complex $(C^{p}_{CE, \mathpzc{int}} (\mathfrak{g}), \delta^p )$ and we define the corresponding \emph{integral} Chevalley-Eilenberg cohomology of the Lie superalgebra $\mathfrak{g}$ in the usual way
\bear \label{ceint}
H^p_{CE, \mathpzc{int}} (\mathfrak{g}) \defeq \frac{\ker \big ( \delta^p : C^p_{CE, \mathpzc{int}} (\mathfrak{g}) \rightarrow C^{p+1}_{CE, \mathpzc{int}} (\mathfrak{g}) \big )}{\mbox{im}  \big ( \delta^{p-1} : C^{p-1}_{CE, \mathpzc{int}} (\mathfrak{g}) \rightarrow C^p_{CE, \mathpzc{int}} (\mathfrak{g}) \big )}.
\eear
Notice that the differential only acts on $S^\bullet \Pi \mathfrak{g}$, as can be seen in \eqref{diff2}. One can therefore alternatively define the above cohomology \eqref{ceint} starting from the cochains $\widehat C^p_{S} (\mathfrak{g}) \defeq S^p \Pi \mathfrak{g}$ on which $\delta$. The related cohomology is then defined as 
\bear
H^p_{S} (\mathfrak{g}) \defeq \frac{\ker \big ( \delta^p : \widehat{C}^p_{S} (\mathfrak{g}) \rightarrow \widehat{C}^{p+1}_{S} (\mathfrak{g}) \big )}{\mbox{im}  \big ( \delta^{p-1} : \widehat{C}^{p-1}_{S} (\mathfrak{g}) \rightarrow \widehat{C}^p_{S} (\mathfrak{g}) \big )},
\eear
and in turn the integral Chevalley-Eilenberg cohomology $H^p_{CE, \mathpzc{int}} (\mathfrak{g})$ becomes a \emph{twist} of $H^p_{S} (\mathfrak{g})$ by $\mbox{Ber}^{\mathpzc{H}} (\mathfrak{g})$, namely
\bear
H^{\bullet}_{CE, \mathpzc{int}} (\mathfrak{g}) \cong \mbox{Ber}^{\mathpzc{H}}(\mathfrak{g}) \otimes H^\bullet_S (\mathfrak{g}).
\eear
Observe that the Haar Berezinian can be seen as a shift by degree $n$ in cohomology and that the cochains are really dual one another, as $C^p_{CE, dif} = S^p \Pi \mathfrak{g}^\ast$ and $C^p_{CE, int} (\mathfrak{g}) = S^p \Pi \mathfrak{g}$. So, the question is: if $\omega \in C^p_{dif} (\mathfrak{g})$ is closed, then is $\omega^\ast \in C^p_{int} (\mathfrak{g})$ closed? And viceversa. This is proved in the following. 

\noindent Let us first consider some calculations in the two formalisms, showing that they are equivalent: consider the case of a $(n-1)$-integral form
\begin{equation}\label{DD}
	\omega^{(n-1)} = \mathpzc{D} \otimes \sum_{A= 1}^{m+n} T^{A} \left( \pi \mathpzc{Y}_{A} \right) \equiv T^{A} \iota_{\mathpzc{Y}_{A}} \omega^{top} \ .
\end{equation}
We can apply the operator $\delta^{(1)} \equiv d$ to $\omega^{(n-1)}$ thus obtaining
\begin{equation}\label{DE}
	\delta^{(1)} \omega^{(n-1)} = \mathpzc{D} \otimes \sum_{B} \sum_A \iota_{(\pi \mathpzc{Y}^\ast_B)} T^{A} \mathcal{L}_{\mathpzc{Y}_B} \left( \pi \mathpzc{Y}_{A} \right) = $$ $$ = \mathpzc{D} \otimes \sum_{B} \sum_{A,C} \iota_{(\pi \mathpzc{Y}^\ast_B)} T^{A} f^C_{B A} \left( \pi \mathpzc{Y}_{C} \right) = \mathpzc{D} \otimes \sum_{B} \sum_{A,C} T^{A} f^C_{B A} \delta_{B C} = 0 \ ,
\end{equation}
where we have used the \eqref{LIEA} for the Lie derivative, $\displaystyle \iota_{(\pi \mathpzc{Y}^\ast_A)} \left( \pi \mathpzc{Y}_{B} \right) = \delta^A_{B}$ and the properties of the structure constants. On the other hand we have
\begin{equation}\label{DH}
	d \omega^{(n-1)} = \frac{1}{2} f^A_{B C} \left( \pi \mathpzc{Y}^{*} \right)^B \left( \pi \mathpzc{Y}^{*} \right)^C \iota_{\mathpzc{Y}_A} T^D \iota_{\mathpzc{Y}_D} \omega^{top} = f^A_{BC } \delta^{~B}_A \delta_D^{~C} 
	T^D \omega^{top} = 0 \ .
\end{equation}
Notice that actually we can use the isomorphism $H^\bullet_{CE} \left( \mathfrak{g} , \mathbb{R} \right) \cong H^\bullet_{dR} \left( G \right)^{G}$ (\emph{i.e.} that the Chevalley-Eilenberg cohomology of the superalgebra $\mathfrak{g}$ is isomorphic to the de Rham cohomology of the supergroup $G$ restricted to the left-invariant forms) to obtain \eqref{DH} in a different way:
\begin{equation}\label{DI}
	d \omega^{(n-1)} = d T^D \iota_{\mathpzc{Y}_D} \omega^{top} = T^D \mathcal{L}_{\mathpzc{Y}_D} \omega^{top} + \left( -1 \right)^{|\pi \mathpzc{Y}_D |} T^D \iota_{\mathpzc{Y}_D} d \omega^{top} = 0 \ ,
\end{equation}
where we have used the fact that $\omega^{top}$ is the Haar Berezinian tensor density in $\mbox{Ber}^\mathpzc{H} (\mathfrak{g})$, hence the (left) invariant top form. The previous example is two-folded: first it is an example of calculation in both realisations with a check of equivalence, second it shows that the Haar Berezinian $\mathpzc{D} \equiv \omega^{top}$, which is obviously closed with respect to $\delta^{(\bullet)} \equiv d$, \emph{is not exact}, thus showing that it is always a cohomology representative.

\subsection{Isomorphism Between Superform and Integral Form Cohomologies.}

\noindent In this section we show that the cohomology of superforms is isomorphic to the cohomology of integral forms. In order to do so, we will use the formalism where the Haar Berezinian is treated as a differential form as in \eqref{BerForm} and the nilpotent operator is actually the Cartan differential. The proof for integral forms written as in \eqref{IntFormPoly} with respect to the differential \eqref{DiffOpeDelta} follows from the \virgolette dictionary" between the two established formalisms.

\noindent Let us start by considering a superform $\omega^{(1)}$, such that $d \omega^{(1)} =0$. We define its \virgolette Berezinian complement" $\star \omega^{(1)}$ as
\bear\label{IBSICB}
\xymatrix@R=1.5pt{
\star : \Omega^1_{CE, \mathpzc{dif}} (\mathfrak{g}) \ar[r] & \Omega^{n-1}_{CE, \mathpzc{int}} (\mathfrak{g}) \\
\omega^{(1)} \ar@{|->}[r] & \star \omega^{(1)} = \left( \star \omega \right)^{(n-1)} \defeq \iota_\mathpzc{Y} \omega^{top}_\mathfrak{g} \ ,
}
\eear
where $\displaystyle \omega^{(1)} (\mathpzc{Y}) = 1$, \emph{i.e.}\ $\pi \mathpzc{Y}$ is the vector field dual to $\omega^{(1)}$. Then we have $\displaystyle d \star \omega^{(1)} = d \iota_\mathpzc{Y} \omega^{top}_\mathfrak{g} = 0$, as we have shown in \eqref{DI}. For a generic $p$-superform the generalization follows by extending \eqref{IBSICB} as
\bear\label{IBSICD}
\xymatrix@R=1.5pt{
\star : \Omega^p_{CE, \mathpzc{dif}} (\mathfrak{g}) \ar[r] & \Omega^{n-p}_{CE, \mathpzc{int}} (\mathfrak{g})\\
\omega^{(p)} \ar@{|->}[r] & \star \omega^{(p)} = \left( \star \omega \right)^{(n-p)} \defeq \iota_{\mathpzc{Y}_1} \ldots \iota_{\mathpzc{Y}_p} \omega^{top}_\mathfrak{g} \ ,
}
\eear
where $\displaystyle \omega^{(p)} (\mathpzc{Y}_1 , \ldots , \mathpzc{Y}_p) = 1 $. Given $\omega^{(p)}\in H^{p}_{CE, \mathpzc{dif}} (\mathfrak{g})$, we have
\begin{equation}\label{IBSICF}
	d \left( \omega_{A_1 \ldots A_p} \left( \pi \mathpzc{Y}^* \right)^{A_1}  \wedge \ldots \wedge \left( \pi \mathpzc{Y}^* \right)^{A_p} \right) = p \omega_{A_1 \ldots A_p} f^{A_1}_{~~R S} \left( \pi \mathpzc{Y}^* \right)^R \left( \pi \mathpzc{Y}^* \right)^S \left( \pi \mathpzc{Y}^* \right)^{A_2} \wedge \ldots \wedge \left( \pi \mathpzc{Y}^* \right)^{A_p} = 0 $$ $$ \iff \ \omega_{A_1 \ldots A_p} f^{A_1}_{~~R S} = 0 \ .
\end{equation}
We now show that this condition implies $d \star \omega^{(p)} = 0$. First of all, we observe that the integral form dual to $\omega^{(p)}$ reads
\begin{equation}\label{IBSICFA}
	\star \omega^{(p)} = T^{A_1 \ldots A_p} \iota_{\mathpzc{Y}^{A_1}} \ldots \iota_{\mathpzc{Y}^{A_p}} \omega^{top}_\mathfrak{g} \ , \ \mbox{ such that } \ T^{A_1 \ldots A_p} \omega_{A_1 \ldots A_p} = 1 \ .
\end{equation}
It is easy to see that
\begin{equation}\label{IBSICG}
	d \star \omega^{(p)}=0 \ \iff \ T^{A_1 A_2 \ldots A_{p}} f_{~~A_1 A_2}^R = 0 \ .
\end{equation}
Recalling that every basic classical Lie superalgebra admits a non-degenerate bilinear form, see \emph{e.g.}\ \cite{Frappat:1996pb}, we can use the (non-degenerate) bilinear form $g_{AB}$ in order to write the coefficients $T$ of the integral form in terms of the coefficients $\omega$ of the superform as
\begin{equation}\label{IBSICH}
	T^{A_1 A_2 \ldots A_{p}} = \frac{1}{||\omega||^2} g^{A_1 B_1} \ldots g^{A_p B_p} \omega_{B_1 \ldots B_p} \ , \ \text{where} \ ||\omega||^2 = \omega_{A_1 \ldots A_p} g^{A_1 B_1} \ldots g^{A_p B_p} \omega_{B_1 \ldots B_p} \ .
\end{equation}
By substituting \eqref{IBSICH} in the left hand side of \eqref{IBSICG}, we obtain
\begin{equation}\label{IBSICI}
	\frac{1}{||\omega||^2}g^{A_1 B_1} \ldots g^{A_p B_p} \omega_{B_1 \ldots B_p} f_{~~A_1 A_2}^R = \frac{1}{||\omega||^2} g^{A_3 B_3} \ldots g^{A_p B_p} \omega_{B_1 \ldots B_p} f^{R B_1 B_2} = 0 \ ,
\end{equation}
as a consequence of \eqref{IBSICF}. Hence the result $d \omega^{(p)} = 0 \ \implies \ d \star \omega^{(p)} = 0$. The converse can be shown in a similar way.

From the previous argument we can now infer the isomorphism between the cohomologies of super and integral forms. In particular, if $\displaystyle \omega^{(p)} \in H^{p}_{CE, \mathpzc{dif}} (\mathfrak{g})$, we have
\begin{equation}
	\omega^{(p)} \wedge \star \omega^{(p)} = \omega^{top}_\mathfrak{g} \in \mbox{Ber}^{\mathpzc{H}}(\mathfrak{g}) \ .
\end{equation}
By contradiction, let us assume $\left( \star \omega \right)^{(n-p)} = d \Lambda^{(n-p-1)}$, we get
\begin{equation}
	\omega^{top}_\mathfrak{g} = d \left( \omega^{(p)} \wedge \Lambda^{(n-p-1)} \right) \ ,
\end{equation}
contradicting that $\omega^{top}_\mathfrak{g}$ is a cohomology representative as shown in the previous section. This argument shows that the operator $\star$ is indeed an isomorphism:
\begin{equation}
	\star : H^{\bullet}_{CE, \mathpzc{dif}} (\mathfrak{g}) \overset{\cong}{\underset{}{\longrightarrow}} H^{n-\bullet}_{CE, \mathpzc{int}} (\mathfrak{g}) \ .
\end{equation}

\section{Poincar\'{e} Polynomials and Betti Numbers}

\noindent Before we move to compute examples of Chevalley-Eilenberg cohomologies, we review the definition of Poincar\'e series and Poincar\'e polynomials. For $X$ a \emph{graded} $k$-vector space with direct decomposition into $p$-degree homogeneous subspaces given by $X = \bigoplus_{p \in \mathbb{Z}} X_p$ we call the formal series 
\begin{eqnarray}
\label{POIA}
\mathpzc{P}_X(t) = \sum_p ({\rm dim}_k \, X_p) (-t)^p
\end{eqnarray}
the \emph{Poincar\'e series} of $X$. Notice that we have implicitly assumed that $X$ is a of \emph{finite type}, \emph{i.e.} its homogeneous subspaces $X_p$ are finite dimensional for every $p.$ The unconventional sign in $(-t)^p$ takes into account the \emph{parity} of $X_p$, which takes values in $\mathbb{Z}_2$ and it is given by $p \, \mbox{mod}\, 2$: this will be particularly useful in the super setting. If also $\dim_k X$ is finite, then $\mathpzc{P}_X(t)$ becomes a polynomial $\mathpzc{P}_X [t]$, called \emph{Poincar\'e polynomial} of $X$. The evaluation of the Poincar\'e polynomial at $t=1$ yields the so-called \emph{Euler characteristics} $\chi_{_X} = \mathpzc{P}_X[t=1] = \sum_p (-1)^p \dim_k X_p$ of $X$. If we assume that the pair $(X, \delta)$ is a differential complex for $X$ a graded vector space and $\delta : X_p \rightarrow X_{p+1}$ for any $p$, then the cohomology $H_{\delta}^\bullet (X) = \bigoplus_{p\in \mathbb{Z}} H_{\delta}^p(X)$ is a graded space. Here we are interested into the case of the de Rham cohomology, where $X = \bigwedge^\bullet \mathcal{T}^{\ast} M$, \emph{i.e.} the exterior bundle of a certain differentiable manifold $M$ and the differential $\delta = d : \bigwedge^p \mathcal{T}^\ast M \rightarrow \bigwedge^{p+1} \mathcal{T} M$ is the de Rham differential: then $H^\bullet_{dR}(M)$ is a graded vector space and we call $b_p (M) \defeq \dim_k H^p_{dR} (M)$ the $p$-th Betti number of $M$. The Poincar\'e polynomial of $M$, defined as (Euler-Poincar\'e formula) 
\bear
\mathpzc{P}_{M} [t] \defeq \mathpzc{P}_{H_{dR} (M)}[t] = \sum_{p} b_p (M) (-t)^p
\eear
is the generating function of the Betti numbers of $M$. This property is known as {\it telescopic nesting} which implies that from the easy computation of $P_X(t)$, one deduces $P_{H(X)}(t)$. From the latter 
one can read the cohomology classes by their gradings and the parity.

Even if the notion of Betti numbers is originally related to the topology of a certain manifold or topological space, by extension, in this paper we will call \emph{Betti numbers} the dimensions of any cohomology space valued in a field, in particular, we will call $p$-th Betti numbers of a certain Lie (super)algebra the dimension of its Chevalley-Eilenberg $p$-cohomology group $b_p (\mathfrak{g}) = \dim_k H^p_{CE} (\mathfrak{g}),$ so that the Poincar\'e series of the Lie (super)algebra $\mathfrak{g}$ is the generating function of its Betti number
\bear
\mathpzc{P}_{\mathfrak{g}} (t) = \sum_p b_p (\mathfrak{g}) (-t)^p.
\eear
Notice that we used the notation $\mathpzc{P} (t)$ on purpose: indeed, as we shall see, $H^\bullet_{CE} (\mathfrak{g})$ is not in general finite dimensional for a generic Lie superalgebra $\mathfrak{g}$.
In this context, we can retrieve some useful results using the Poincar\'e series. For example, K\"{u}nneth theorem, which computes the cohomology of products of spaces, can simply be written as
\bear
\mathpzc{P}_{X \otimes Y} (t) = \mathpzc{P}_X(t) \cdot \mathpzc{P}_{Y}(t).
\eear

Sometimes, it is useful to introduce a second grading. In that case the space is said to be 
{\it bigraded vector space} $X = \sum_{p,q \in \mathbb{Z}} X^{p,q}$, then the gradation 
$X = \sum_r X^{r}$ given by 
\begin{eqnarray}
\label{POID}
X^r = \sum_{p+q=r} X^{p,q}
\end{eqnarray}
is called the {\it induced total gradation}. One can write a {\it double Poincar\'e series} 
\begin{eqnarray}
\label{POIE}
\mathpzc{P}_X(t,s) = \sum_{p,q} (-t)^p s^q {\rm dim}X^{p,q}
\end{eqnarray}
 which, in any case, allows an easier identification of cohomological classes (see, e.g., \cite{Grassi:2005jz} where double Poincar\'e series have been used to select different type cohomologies).

\section{Chevalley-Eilenberg Cohomology: Computations}

\subsection{Dimension 1: Example of Infinite Cohomology}

\noindent In order to get familiar with cohomology computations of Lie superalgebras we start from a \virgolette simple model'', that is the Lie superalgebra of the \emph{supertranslations} of the superspace $\mathbb{R}^{1|2}$, which we will denote $\mathpzc{susy} (\mathbb{R}^{1|2})$, and we spell out all the details. Starting from the supermanifold structure, here - and in the following examples - the superspace $\mathbb{R}^{1|2}$ is actually not to be looked at as just the \virgolette bare'' flat superspace $\mathbb{R}^{1|2}$, characterized by the pair $(\mathbb{R}, \mathcal{O}_{\mathbb{R}} \otimes \wedge^\bullet [ \theta_1, \theta_2 ])$ as a ringed space, where the first entry is just the ordinary manifold $\mathbb{R}$ and the second entry is a sheaf of exterior algebras generated over two anti-commuting variables $\theta_1 $ and $\theta_2$, \emph{i.e.}\ the structure sheaf $\mathcal{O}_{\mathbb{R}^{1|2}}$ of the supermanifold $\mathbb{R}^{1|2}$. Instead, $\mathbb{R}^{1|2}$ carries some additional data, namely an \emph{odd distribution} of the tangent bundle of $\mathbb{R}^{1|2}$, we denote it as $\mathpzc{Susy} \subset \mathcal{T} (\mathbb{R}^{1|2})$, which is generated by the fields
\bear
\mathcal{Q}_1 \defeq \frac{\partial}{\partial \theta_1} - \theta_2 \frac{\partial}{\partial x}, \qquad \mathcal{Q}_2 \defeq \frac{\partial}{\partial {\theta_2}} - \theta_1 \frac{\partial}{\partial x},
\eear
and which satisfies the commutation relations 
\bear \label{susy1}
\{\mathcal{Q}_\alpha, \mathcal{Q}_\beta \} = 2\delta_{\alpha \beta} \mathcal{P},
\eear
for $\alpha , \beta = 1, 2$, where we have defined $\mathcal{P} \defeq -\frac{\partial}{\partial x}$. This means that the distribution $\mathpzc{Susy}$ generated by $\{ \mathcal{Q}_1, \mathcal{Q}_2\}$ is \emph{non-integrable} and the triple $\{ \mathcal{Q}_1, \mathcal{Q}_2, \mathcal{P} \}$ generates the tangent bundle at any point.
Adding to the previous relations \eqref{susy1} also the obvious commutation relations $[\mathcal{P}, \mathcal{P}] = 0$ and $[\mathcal{P}, \mathcal{Q}_i] = 0$ for any $i=1,2$ one gets the \emph{supersymmetry translation algebra}, or supertranslation algebra for short, which we denote $\mathpzc{susy} (\mathbb{R}^{1|2})$.\\
Switching from fields to forms, in order to write the cochains $C^p_{CE} (\mathpzc{susy} (\mathbb{R}^{1|2}) ) =  S^p \Pi  \mathpzc{susy} (\mathbb{R}^{1|2})^\ast,$ we have to find the dual vielbeins (up to a parity shift) to the above fields. These are 
\bear
\mathpzc{V} \defeq dx - \theta^1 d\theta^2 - \theta^2 d\theta^1, \quad \psi^\alpha = d\theta^\alpha, 
\eear
for $\alpha = 1, 2,$ and it can be easily checked that $\mathpzc{V} (\pi \mathcal{P}) = 1 $ and $ \psi^\alpha (\pi \mathcal{Q}_\beta ) = \delta^\alpha_\beta$. The Maurer-Cartan equations for the vielbeins $C^1_{CE} ( \mathpzc{susy}( \mathbb{R}^{1|2})) = \{\psi^\alpha | \mathpzc{V} \}$ are easily computed to be
\bear
d \mathpzc{V} =  -2\psi^1 \psi^2, \qquad d \psi^\alpha = 0,
\eear
for $\alpha = 1, 2.$ Now the cohomology is readily computed observing that terms involving $\mathpzc{V}$ will never be closed and terms involving the product $\psi^1 \psi^2$ will always be exact. This leads to the following \emph{differential} Chevalley-Eilenberg cohomology
\bear
H^p_{CE, \mathpzc{dif}} (\mathpzc{susy} (\mathbb{R}^{1|2})) \cong \mathbb{R}\cdot \{ (\psi^\alpha )^p \}, 
\eear
for any $p \geq 0$ and $\alpha = 1, 2.$ Assigning the weights to Maurer-Cartan forms according to $\mathpzc{W} (\mathpzc{V}) = 1$ and $\mathpzc{W} (\psi^\alpha ) = 1/2$ for any $\alpha = 1, 2$, one finds for the Poincar\'e series
\bear
\mathpzc{P}^{\mathpzc{dif}}_{\mathpzc{susy} (\mathbb{R}^{1|2})} (t) = \frac{1-t}{\left( 1- \sqrt{t} \right)^2} = \frac{1+ \sqrt{t}}{1-\sqrt{t}} = 1 + 2 \sum_{n=1}^\infty t^{{n}/{2}} \ ,
\eear 
where the denominator has been expanded around 0. This is in agreement with the previous computation, which indeed says that the \emph{Betti numbers} of the superalgebra are 
\bear
b_1 (\mathpzc{susy} (\mathbb{R}^{1|2})= 1, \quad b_{p\geq 1} (\mathpzc{susy} (\mathbb{R}^{1|2}) = 2.
\eear
Let us now look at the integral Chevalley-Eilenberg cohomology. Repeating the above analysis, posing 
\bear
\mathpzc{D}_{\mathpzc{susy}(\mathbb{R}^{1|2})} \defeq \mathpzc{V} %\sum_{\alpha, \beta = 1}^2
 \epsilon_{\alpha \beta} \delta (\psi^\alpha)\delta (\psi^\beta) \in \mbox{Ber}^{\mathpzc{H}} (\mathpzc{susy} (\mathbb{R}^{1|2})),
\eear
one finds that 
\bear
H^{1-p}_{CE, \mathpzc{int}} (\mathpzc{susy} (\mathbb{R}^{1|2})) = \mathbb{R} \cdot \left \{ (\iota_{\pi \mathcal{Q}_1 } )^p \mathpzc{D}_{\mathpzc{susy} (\mathbb{R}^{1|2})} , \; (\iota_{\pi \mathcal{Q}_2  } )^p \mathpzc{D}_{\mathpzc{susy}( \mathbb{R}^{1|2})}\right  \}, 
\eear
where we notice in particular that the (Haar) Berezinian $(\mathpzc{D}_{\mathpzc{susy} (\mathbb{R}^{1|2})})$ generates the integral 1-cohomology $H^1_{CE, \mathpzc{int}} (\mathpzc{susy} (\mathbb{R}^{1|2})).$ Mirroring what above, terms coming from a double contraction $\iota_{\pi \mathcal{Q}_1} \iota_{\pi \mathcal{Q}_2} $ are not closed, while terms that do not contain $\mathpzc{V}$ are exact. Just like above, this matches the Poincar\'e series computation, namely 
\bear
\mathpzc{P}^{\mathpzc{int}}_{\mathpzc{susy} (\mathbb{R}^{1|2})} (t) = \frac{1-t}{\left( 1- {1}/{\sqrt{t}} \right)^2} \left( \frac{-1}{\sqrt{t}} \right)^2 = \frac{1+ \sqrt{t}}{1-\sqrt{t}} = -1 - 2 \sum_{n=1}^\infty t^{-{n}/{2}},
\eear
where now we expanded the denominator around infinity, as to represent the cohomology spaces with negative form degree.

\noindent Before we move to higher dimensional examples, some remarks are in order. First of all, already in this example it has to be noted an obvious yet striking difference between the ordinary and the super Chevalley-Eilenberg cohomology, namely the fact that even the cohomology of finite dimensional Lie superalgebras can be \emph{infinite}, whereas clearly every finite-dimensional Lie algebras has a finite-dimensional Chevalley-Eilenberg cohomology. This is allowed by the very structure of the cochain complex, which is not bounded from above for a true Lie superalgebra, \emph{i.e.}\ a Lie superalgebra whose odd dimension is different from zero, and by the structure of the commutators, which can leave \virgolette unconstrained'' an even form, such as in the case of $\psi^\alpha$ above.

\subsection{Dimension 2: \virgolette Flat'' and \virgolette Curved'' Cases}

\noindent We now pass to study some more interesting cases of cohomology of Lie superalgebras, which both have two bosonic dimensions. Namely we study the Lie superalgebra of supertranslations related to the superspace $\mathbb{R}^{1,1 | 2}$, which we will call \emph{flat} superspace as it is constructed over the Minkowski space $\mathbb{R}^{1,1}$, and the Lie superalgebra $\mathfrak{u} (1|1)$. For the sake of completeness and readability of the paper, the general mathematical structure of the Lie superalgebra $\mathfrak{u} (n|m)$ is described in Appendix \ref{unit}.

\subsubsection{Flat Case: Supertranslations of the ${D=2}$, ${\mathcal{N}=1}$ Superspace}

\noindent Repeating the above discussion for the $1$-dimensional case, one is lead to consider the algebra of supertranslations generated by the following vector fields
\begin{align}
\mathcal{Q}_\alpha \defeq \frac{\partial}{\partial \theta^\alpha} - %\sum_{i=0}^1 \sum_{\beta = 1 }^2
 (\theta^{\beta} \Gamma_{\beta \alpha})^i \frac{\partial}{\partial x^i}, \qquad \mathcal{P}_i \defeq - \frac{\partial}{\partial x^i},   
\end{align}
where $x^i | \theta^\alpha$ for $i = 0,1$ and $\alpha = 1, 2$ are coordinate for $\mathbb{R}^{1,1 |2}$ and the gamma matrices $\Gamma_{\alpha \beta}^i$ generating the {spin representation} of $\mathfrak{so}(1,1)$ in which the odd coordinates transform, are given by
\begin{equation}
	\Gamma^1_{\alpha \beta} = \left( \begin{array}{cc}
		1 & 0 \\
		0 & -1
	\end{array} \right), \qquad  \Gamma^2_{\alpha \beta} = \left ( \begin{array}{cc}
		0 & 1 \\
		1 & 0
	\end{array} \right).
\end{equation}
The non-trivial commutation relations (supersymmetries) characterizing the Lie superalgebra $\mathpzc{susy} (\mathbb{R}^{1,1|2})$ read
\bear
\{\mathcal{Q}_\alpha, \mathcal{Q}_{\beta} \} = - 2 %\sum_{i=0}^1
 \Gamma^i_{\alpha \beta} \mathcal{P}_{i}.
\eear
Dually, we introduce the vielbeins, which will generate the cochains for the Lie superalgebra. These are given by
\bear
\mathpzc{V}^i = dx^a - %\sum_{\alpha, \beta = 1 }^2
 \theta^\alpha \Gamma^a_{\alpha \beta} d \theta^\beta, \quad \psi^\alpha = d \theta^\alpha 
\eear
again for $i = 0,1$ and $\alpha = 1,2$. The Maurer-Cartan equations reads
\bear
d\mathpzc{V}^i = %\sum_{\alpha, \beta}
 \psi^\alpha \Gamma^a_{\alpha \beta} \psi^\beta, \qquad d \psi^\alpha = 0, 
\eear
leading to the following cohomology:
\begin{align}
& H^0_{CE, \mathpzc{dif}} (\mathpzc{susy} (\mathbb{R}^{1,1|2})) \cong \mathbb{R} \cdot 1, \qquad H^1_{CE, \mathpzc{dif}} (\mathpzc{susy} (\mathbb{R}^{1,1|2})) \cong \mathbb{R} \cdot \{ \psi^1, \psi^2 \} \nonumber \\
& H^2_{CE, \mathpzc{dif}} (\mathpzc{susy} (\mathbb{R}^{1,1|2})) \cong \mathbb{R} \cdot \bigg  \{ \sum_{\alpha = 1}^2(\psi^\alpha)^2 \bigg \}, \qquad H^{p > 2}_{CE, \mathpzc{dif}} (\mathpzc{susy} (\mathbb{R}^{1,1|2})) \cong 0. 
\end{align}
This result is in agreement with the Poincar\'e polynomial, as indeed
\bear
\mathpzc{P}^{\mathpzc{dif}}_{\mathpzc{susy} (\mathbb{R}^{1,1|2})} [\sqrt{t}] = \frac{\left( 1-t \right)^2}{\left( 1- \sqrt{t} \right)^2} = 1 + 2 \sqrt{t} + t.
\eear
Switching to integral forms, posing as above 
\bear
\mathpzc{D}_{\mathpzc{susy}(\mathbb{R}^{1,1|2})} \defeq \mathpzc{V}^1 \mathpzc{V}^2 \delta (\psi^1)\delta (\psi^2) \in \mbox{Ber}^{\mathpzc{H}} (\mathpzc{susy} (\mathbb{R}^{1,1|2}),
\eear
and repeating the above computations one gets accordingly that 
\begin{align}
& H^2_{CE, \mathpzc{int}} (\mathpzc{susy} (\mathbb{R}^{1,1|2})) \cong \mathbb{R} \cdot \mathpzc{D}_{\mathpzc{susy}(\mathbb{R}^{1,1|2})}, \qquad H^1_{CE, \mathpzc{dif}} (\mathpzc{susy} (\mathbb{R}^{1,1|2})) \cong \mathbb{R} \cdot \{ \iota_{\pi \mathcal{Q}_\alpha} \mathpzc{D}_{\mathpzc{susy}(\mathbb{R}^{1,1|2})} \} \nonumber \\
& H^0_{CE, \mathpzc{int}} (\mathpzc{susy} (\mathbb{R}^{1,1|2})) \cong \mathbb{R} \cdot \bigg  \{ \sum_{\alpha = 1}^2 (\iota_{\pi \mathcal{Q}_\alpha})^2 \mathpzc{D}_{\mathpzc{susy}(\mathbb{R}^{1,1|2})} \bigg \}, \qquad H^{p < 0}_{CE, \mathpzc{dif}} (\mathpzc{susy} (\mathbb{R}^{1,1|2})) \cong 0. 
\end{align}
The Poincar\'e polynomial reads
\bear
\mathpzc{P}^{\mathpzc{int}}_{\mathpzc{susy} (\mathbb{R}^{1,1|2})} [t]= \frac{\left( 1-t \right)^2}{\left( 1- {1}/{\sqrt{t}} \right)^2} \left( \frac{-1}{\sqrt{t}} \right)^2 = 1 + 2 \sqrt{t} + t.
\eear

\subsubsection{Curved Case: Lie Superalgebra $\mathfrak{u} (1|1)$}

\noindent We now aim at computing the cohomology of the $2|2$-dimensional Lie superalgebra $\mathfrak{u} (1|1)$. Further, later on, we briefly comment on \emph{Cartan's theorem} on the cohomology of \emph{compact} and \emph{connected} Lie groups in the supersetting. Before we start, we recall that, for the sake of the readability of the paper, the construction of the Lie superalgebra $\mathfrak{u}(n|m)$ for arbitrary values of $n$ and $m$ is given in Appendix \ref{unit}.

%\subsubsection{Cohomology of $\mathfrak{u} (1|1)$}

\noindent In the easiest case $\mathfrak{u} (1|1)$, one has a $2|2$-dimensional Lie superalgebra, whose general element can be given in the following form
\bear
X = \left ( \begin{array}{c|c}
i a & \theta + i \psi \\
\hline 
- \psi - i \theta & i b 
\end{array}
\right ),
\eear
for $a, b \in \mathbb{R}$ and $\theta , \psi \in \Pi\mathbb{R},$ so that the even and odd generators can be chosen to be the matrices 
\begin{align}
X_1 = \left ( \begin{array}{c|c}
i  &  \\
\hline 
 &  0
\end{array}
\right ), \quad 
X_2 = \left ( \begin{array}{c|c}
0  &  \\
\hline 
 & i 
\end{array}
\right ), \quad 
\Psi_1 = \left ( \begin{array}{c|c}
  & 1 \\
\hline 
-i &  
\end{array}
\right ), \quad
\Psi_2 = \left ( \begin{array}{c|c}
  & i \\
\hline 
-1 &  
\end{array}
\right ),
\end{align}
together with the commutation relations
\begin{align} \label{cr1}
[X_i, X_j] = 0, \quad [X_1, \Psi_1 ] = \Psi_2,  \quad [X_1, \Psi_2] = - \Psi_1, \quad [X_2, \Psi_1] = - \Psi_2, \quad [X_2, \Psi_2] = \Psi_1  \\
\{ \Psi_1, \Psi_1 \} = -2 X_1 - 2X_2, \quad  \{ \Psi_2, \Psi_2 \} = - 2 X_1 - 2X_2, \quad \{ \Psi_1, \Psi_2 \} = 0 \label{cr2}.
\end{align}
Introducing the dual (up to parity) basis of Maurer-Cartan forms of $\Pi \mathfrak{u} (1|1)^\ast$, defined so that $\Pi \mathfrak{u} (1|1)^\ast = \mbox{Span}_\mathbb{R} \{ \mathpzc{V}^i | \psi^\alpha \}$ for $i = 1, 2$ and $\alpha = 1, 2$, with $V^i (\pi X_j) = \delta^i_j$ and $\psi^\alpha (\pi \Psi_\beta) = \delta^\alpha_\beta$, one sees from \eqref{cr1} and \eqref{cr2} that the Maurer-Cartan equations read
\bear
dV^1 = dV^2 = - \sum_{\alpha =1}^2 (\psi^\alpha)^2, \quad d\psi^1 = \psi^2 \frac{( V^1 - V^2 )}{2}, \quad d \psi^2 = \psi^1  \frac{(- V^1 + V^2 )}{2}.
\eear
Changing the basis to $U \defeq \frac{V^1 - V^2}{2}$ and $W \defeq \frac{V^1 + V^2}{2}$, the Maurer-Cartan equations simplify to
\bear \label{mcUW}
dU = 0, \qquad dW = - \sum_{\alpha =1}^2 (\psi^\alpha)^2, \qquad d\psi^1 = U \psi^2, \qquad d \psi^2 = -U \psi^1.
\eear
Starting from the $p$-cochains $C^p (\mathfrak{u} (1|1)) = S^p (\Pi \mathfrak{u} (1|1)^\ast )$ and using the above Maurer-Cartan equations \eqref{mcUW}, it is not hard to compute the related Chevalley-Eilenberg cohomology:
\bear
H^0_{CE, \mathpzc{dif}} (\mathfrak{u} (1|1)) \cong \mathbb{R} \cdot 1, \qquad H^1_{CE, \mathpzc{dif}} (\mathfrak{u} (1|1)) \cong \mathbb{R} \cdot \{ U \}, \quad H^{p > 1}_{CE, \mathpzc{dif}} (\mathfrak{u} (1|1)) = 0. 
\eear
This is in agreement with the computation of the Poincar\'e polynomial, which is the one of the ordinary 1-dimensional unitary Lie algebra $\mathfrak{u} (1)$ as shown also by Fuks:
\bear
\mathpzc{P}^{\mathpzc{dif}}_{\mathfrak{u} (1|1)} [t] = \mathpzc{P}_{\mathfrak{u} (1)}^{\mathpzc{dif}} [t] = 1-t.
\eear
In the case of integral Chevalley-Eilenberg cohomology one has the following cohomology
\begin{align}
H^2_{CE, \mathpzc{int}} (\mathfrak{u} (1|1)) \cong \mathbb{R} \cdot \mathpzc{D}_{\mathfrak{u}(1|1)}, & \qquad H^1_{CE, \mathpzc{int}} (\mathfrak{u} (1|1)) \cong \mathbb{R} \cdot \{ \iota_{\pi U^\ast} \mathpzc{D}_{\mathfrak{u} (1|1)} \}, \nonumber  \\ & H^{p < 1}_{CE, \mathpzc{int}} (\mathfrak{u} (1|1)) = 0. 
\end{align}
where we have posed again $\mathpzc{D}_{\mathfrak{u} (1|1)} = UW \delta (\psi^1) \delta (\psi^2)$ so that for example the representative of the 1-cohomology group is given by $ W \delta( \psi ) \delta (\psi )$, and accordingly the Poincar\'{e} polynomial is computed to be 
\bear
\mathpzc{P}^{\mathpzc{int}}_{\mathfrak{u} (1|1)} [t] = t^2 - t.
\eear

\subsubsection{A Remark on Cartan Theorem on Compact Lie Groups} 

A crucial result in Lie algebra cohomology theory is a theorem due to Cartan, which states that under the topological assumptions of \emph{compactness} and \emph{connectedness}, the de Rham cohomology of a Lie group $G$ is isomorphic to the cohomology of its Lie algebra (valued in the real numbers), \emph{i.e.}\ $H^p_{dR} (G) \cong H^p_{CE} (\mathfrak{g})$; clearly, the result is remarkable not only from a conceptual point of view, but also from a computational point of view, for it allows to get topological informations on large interesting classes of Lie groups via linear algebra. The above result on $\mathfrak{u} (1|1)$ shows that the result does not hold true in the supersetting, whereas one naively substitutes the ordinary compact Lie group $G$ with a compact Lie supergroup $\mathpzc{G}$ and the Lie algebra $\mathfrak{g}$ with its Lie superalgebra. \\
\noindent Let us look indeed at the Lie supergroup $U (1|1) $ related to $\mathfrak{u}(1|1)$. Especially in this context, it is convenient to introduce the unitary supergroup $U(1|1)$ as the \emph{super Harish-Chandra pair} $(U(1) \times U(1), \mathfrak{u} (1|1)),$ since the categories of Lie supergroups and super Harish-Chandra pairs are indeed equivalent \cite{CCF}. %{\bf forse ci vuole una spiegazione? Facciamo finta di no}. 
As it is well-known \cite{CGM}, the de Rham cohomology of a supermanifold only depends on its underlying topological space, and as such it is completely determined by the first entry, \emph{i.e.}\ the ordinary Lie group, of the super Harish-Chandra pair. In our case, we obtain the cohomology of a 2-torus $S^1 \times S^1 \cong U(1) \times U(1)$:
\bear
H^p_{dR} (U(1|1)) \cong \left \{
\begin{array}{lc}
\mathbb{R} & p = 0\\
\Pi \mathbb{R}^2 & p=1\\
\mathbb{R} & p=2.
\end{array}
\right.
\eear 
This shows that the de Rham cohomology of \emph{compact} Lie supergroups, such as for example $U(1|1)$ which is topologically a 2-torus, is not isomorphic to the Chevalley-Eilenberg cohomology of superforms of their related Lie superalgebras. 

Notice by the way, that the isomorphism is restored once one reduces to deal with the \emph{even} - or \emph{topological} - part of a Lie superalgebra. In other words, if, as a vector space, a Lie superalgebra is such that $\mathfrak{g} = \mathfrak{g}_0 \oplus \mathfrak{g}_1$, and its related (\emph{e.g.}\ via Harish-Chandra pair) Lie supergroup $\mathpzc{G}$ is topologically compact as a (super)manifold, then one finds that for any $p$
\bear 
H^p_{dR} (\mathpzc{G}) \cong H^p_{CE, {\mathpzc{dif}}} (\mathfrak{g}_0 ).
\eear This is readily seen in the above case for the Lie superalgebra $\mathfrak{u} (1|1)$, where modding out the odd part of the underlying vector space, one is left with Maurer-Cartan equations of the form $dU=0$ and $dW=0$, which indeed lead to the same cohomology of the $2$-torus. \\
Once again, it has therefore to be stressed that whilst fermions play really no role when computing de Rham cohomology of a supermanifold as nilpotents do not modify topology, in the case of Chevalley-Eilenberg cohomology of a Lie superalgebra, which is ultimately determined by the structure of commutators or, equivalently, by the Maurer-Cartan equations, fermions play a crucial role and they do indeed determine the cohomology structure, which might be very different - either richer or poorer -  from the cohomology of the topological even part of the superalgebra.
%There is a further option, the complete cohomology has to be searched in the complete complex of forms
%\cite{CCGN2}. 

\subsection{Dimension 3: \virgolette Flat'' and \virgolette Curved'' Cases}

\noindent In this section we study two examples of superalgebras that have 3 bosonic dimensions. In particular, we study the cohomology of the Lie superalgebra of superstranslations of flat superspace $\mathbb{R}^{1,2|2}$ and, after having reviewed (in appendix B) the construction of the (simple) Lie superalgebra $\mathfrak{osp} (n|2m)$ for generic values of $n$ and $m$ we study the cohomology of its simplest case, namely $\mathfrak{osp}(1|2),$ corresponding to the \emph{classical simple} Lie superalgebra $B(0,1)$ in Kac's classification. 

\subsubsection{Flat Case: Superstranslations of $D=3$, $\mathcal{N}=1$ Superspace}

\noindent We describe the supermanifold $\mathbb{R}^{1,2|2}$, based on the Minkowski space $\mathbb{R}^{1,2}$ by a set of two coordinates $(x^{a}%
,\theta^{\a})$. In terms of these coordinates, we have the following supersymmetry generators 
\bear \label{3dimgen}
\mathcal{Q}_\alpha \defeq \frac{\partial}{\partial \theta^{\a}} - %\sum_{a=0}^2 \sum_{\beta = 1}^2 
(\gamma_{\alpha \beta} \theta^\beta)^a \frac{\partial}{\partial x^{a}}.
\eear
Notice that in the above we are using \emph{real} and \emph{symmetric} gamma matrices $\gamma^a_{\a \b}$, which are defined via \emph{charge conjugation}, given by the Pauli matrix $C \defeq - i \sigma_2 = \epsilon_{\a \b}$, so that we have
\bear  
\label{dicA}
&&\gamma^0_{\a\b} \defeq (C \Gamma^0)_{\alpha \beta} = - {\mathbf 1}, \qquad \gamma^1_{\a\b} \defeq (C \Gamma^1)_{\alpha \beta} = \sigma^3, \qquad \gamma^2_{\a\b} \defeq (C \Gamma^2)_{\alpha \beta} = - \sigma^1\,,
%C_{\a\b} = i \sigma^2 = \epsilon_{\a\b}\,. 
\eear
where\begin{eqnarray}
\Gamma^0 \defeq i \sigma^2 = \left ( \begin{array}{cc}
0 & 1 \\
-1 & 0
\end{array}
\right ), \qquad
\Gamma^1 \defeq i \sigma^1 = \left ( \begin{array}{cc}
0 & 1 \\
1 & 0
\end{array}
\right ), \qquad
\Gamma^2 \defeq  \sigma^3 = \left ( \begin{array}{cc}
1 & 0 \\
0 & -1
\end{array}
\right ),
\end{eqnarray} 
which satisfies the the Clifford algebra relation $\{ \Gamma^a, \Gamma^b \} = 2 \eta^{ab} \mathbf{1}$, with $\eta^{ab}$ the Minkowski metric and they must be looked at as $(\Gamma^a)^{\alpha}_{\; \beta}$ from the point of view of spinor indices. Defining as above $\mathcal{P}_a = - \frac{\partial}{\partial x^a}$, we have the commutation relations of the algebra $\mathpzc{susy} (\mathbb{R}^{1,2|2})$
\bear
\{ \mathcal{Q}_\alpha, \mathcal{Q}_{\beta} \} =  2 %\sum_{i = 0}^2 
\gamma^a_{\alpha \beta} \mathcal{P}_a.
\eear
Switching to forms, we have the following dual (up to parity) basis of Maurer-Cartan $1$-forms $C^1_{CE, \mathpzc{dif}} (\mathpzc{susy} (\mathbb{R}^{1,2|2})) \defeq \{ \psi^\alpha | \mathpzc{V}^a \} $ for $a = 0,\ldots, 2$ and $\alpha = 1, 2$ with
\bear
\mathpzc{V}^a \defeq dx^a -  \theta^\a \gamma^a_{\a \b} d\theta^\b , \qquad \psi^\a \defeq d\theta^\a.
\eear
From the commutation relation above one reads the Maurer-Cartan equations
\bear
d \mathpzc{V}^a =  \psi^\a \gamma^a_{\a \b} \psi^\b, \qquad d \psi^\a = 0\,, 
\eear
which in turns leads to the following (differential) Chevalley-Eilenberg cohomology\footnote{Some results have been already appeared in the liturature \cite{Brandt:2010fa}.}
\begin{align}
&H^0_{CE, \mathpzc{dif}}(\mathpzc{susy} (\mathbb{R}^{1,2|2} ))  \cong  \mathbb{R} \cdot 1, \nonumber \\
&H^1_{CE, \mathpzc{dif}} (\mathpzc{susy} (\mathbb{R}^{1,2|2}) ) \cong \mathbb{R} \cdot \{ \psi^\alpha \}, \nonumber \\
& H^2_{CE, \mathpzc{dif}} (\mathpzc{susy} ( \mathbb{R}^{1,2|2} ) ) \cong \mathbb{R} \cdot \bigg \{  \mathpzc{V}^a \gamma_{a, \alpha \beta } \psi^\beta \bigg \},  \nonumber \\
& H^3_{CE, \mathpzc{dif}} (\mathpzc{susy} ( \mathbb{R}^{1,2|2} ) ) \cong \mathbb{R} \cdot \bigg \{ \mathpzc{V}^a \psi^\alpha \gamma_{a, \alpha \beta } \psi^\beta \bigg \}, 
\end{align}
and $H^{p>3}_{CE, \mathpzc{dif}} (\mathpzc{susy} ( \mathbb{R}^{1,2|2} ) ) = 0.$ The closure of the above $2$-forms and $3$-form is easily seen by using Fierz identities. Accordingly, the computations of the Poincar\'e polynomial in the present case gives 
\begin{eqnarray}
\label{PPB}
\mathpzc{P}_{\mathpzc{susy} (\mathbb{R}^{1,2|2})}^{\mathpzc{dif}} [\sqrt{t}] = \frac{(1-t)^3}{(1-\sqrt{t})^2}=  (1-\sqrt{t}) (1+ \sqrt{t})^3 = 1 + 2 \sqrt{t} - 2 t \sqrt{t}  - t^2,
\end{eqnarray}
where we observe that the signs indeed match the parity of the representatives. Passing to the integral Chevalley-Eilenberg cohomology we find, keeping explicit the structure of the generators and leaving the wedge product understood,
\begin{align}
& H^3_{CE, \mathpzc{int}} (\mathpzc{susy} (\mathbb{R}^{1,2|2})) \cong \mathbb{R} \cdot \{ \mathpzc{V}^0 \mathpzc{V}^1 \mathpzc{V}^2 \delta (\psi^1)\delta (\psi^2) \mathpzc{}\}, \nonumber \\
& H^2_{CE, \mathpzc{dif}} (\mathpzc{susy} (\mathbb{R}^{1,1|2})) \cong \mathbb{R} \cdot \{ \mathpzc{V}^0 \mathpzc{V}^1 \mathpzc{V}^2 \iota_{\pi \mathcal{Q}_{\alpha}} \delta (\psi^1) \delta (\psi^2)  \} \nonumber \\
& H^1_{CE, \mathpzc{int}} (\mathpzc{susy} (\mathbb{R}^{1,1|2})) \cong \mathbb{R} \cdot  \bigg \{  \mathpzc{V}^a \mathpzc{V}^b \gamma_{ab, \a \b} \iota_{\pi \mathcal{Q}_\b} \delta (\psi^1) \delta (\psi^2)  \bigg \},  \nonumber \\
& H^{0}_{CE, \mathpzc{int}} (\mathpzc{susy} (\mathbb{R}^{1,1|2})) \cong \mathbb{R} \cdot  \bigg \{ \mathpzc{V}^a \mathpzc{V}^b \iota_{\pi \mathcal{Q}_{\a} }\gamma_{ab,\a \b} \iota_{\pi \mathcal{Q}_\b} \delta (\psi^1) \delta (\psi^2)  \bigg \} \nonumber \\
& H^{p < 0}_{CE, \mathpzc{dif}} (\mathpzc{susy} (\mathbb{R}^{1,1|2})) \cong 0, 
\end{align}
where we have defined $\gamma^{ab} \defeq \frac{1}{2}[\gamma^a, \gamma^b]$ and it can be seen that $\gamma^{ab}_{\a\b} =  \epsilon^{ab}_c \gamma_{\a\b}^c$. Also notice that, as above, the highest integral cohomology group is indeed generated by the (Haar) Berezinian, namely $\mathbb{R} \cdot \mathpzc{D}_{\mathpzc{susy}(\mathbb{R}^{1,1|2})} = \epsilon_{abc}\mathpzc{V}^a \mathpzc{V}^b \mathpzc{V}^c \epsilon_{\a \b}\delta (\psi^\a)\delta (\psi^\b).$ The Poincar\'e polynomial reads
\bear
\mathpzc{P}^{\mathpzc{int}}_{\mathpzc{susy} (\mathbb{R}^{1,2|2})} [\sqrt{t}]= \frac{\left( 1-t \right)^3}{\left( 1- {1}/{\sqrt{t}} \right)^2} \left( \frac{-1}{\sqrt{t}} \right)^2 = 1 + 2 \sqrt{t} - 2 t \sqrt{t} - t^2.
\eear
where we have used the assignement of the charges as in the previous sections. The factor $(1-t)^3$ is due to 
$\mathpzc{V}^a$'s, the factor $(1- 1/\sqrt{t})^2$ in the denominator is due to the contractions 
$\iota_{\pi \mathcal{Q}_\a}$ (being a contraction w.r.t. an odd vector a commuting object). The factor $(-1/\sqrt{t})^2$ 
is due to the term $\delta (\psi^\a)\delta (\psi^\b)$. Notice that the Poincar\'e polynomial 
is exactly the same as in (\ref{PPB}).  \\

As it is known, beside integral and differential forms, there are also forms with non-maximal and non-zero picture number, which are usually called \emph{pseudoforms} \cite{Witten,CatenacciGrassiNoja2,CGN,CCGN,CA1,CA2}. Just as a hint, in the present case, pseudoforms have picture number 1 and form other two 
complexes unbounded both from above and from below. The prototype for these forms is 
$\mathpzc{V} \dots \mathpzc{V} (\psi^1)^a (\iota_2)^b \delta(\psi^2)$ where $a, b \geq 0$ (by exchanging $\psi^1$ with $\psi^2$, we get the 
other complex).  Counting again the scaling dimensions we have 
\begin{eqnarray}
\label{PPE}
\mathpzc{P}_{\rm pseudo}[\sqrt{t}] = \frac{(1-t)^3}{\left(1-\frac{1}{\sqrt{t}}\right) (1 - \sqrt{t})} \left(\frac{-1}{\sqrt{t}}\right)=1 + 2 \sqrt{t} - 2 t \sqrt{t}  - t^2
\end{eqnarray}
where the factor $(1-t)^3$ is due to the $\mathpzc{V}$'s, $1/(1-\frac{1}{\sqrt{t}})$ takes into account 
the powers of $\iota_2$, $1/(1-\sqrt{t})$ takes into account  the powers of $\psi^1$. Finally, $ \frac{-1}{\sqrt{t}}$ 
represents $\delta(\psi^2)$ which scales as $1/\sqrt{t}$ and the minus sign takes into account 
the fermion nature of a single delta. We do not explore any further this \virgolette sector'' of the cohomology,  but 
it will turn out to be crucial for a complete understanding of the Chevalley-Eilenberg cohomology in this extended framework \cite{CCGN2}.

\subsubsection{Curved Case: Lie Superalgebra $\mathfrak{osp} (1|2)$ and its \.{I}n\"on\"u-Wigner Contraction to $\mathpzc{susy}({\mathbb{R}^{1,2|2}})$}

\noindent For the sake of readability of the paper, we review in Appendix \ref{appB} the construction of the orthosymplectic Lie superalgebra $\mathfrak{osp}(n|2m)$ for generic values of $n$ and $m$. Here we restrict to the case $\mathfrak{osp}(1|2) = B(0,1)$ and compute its cohomology. Last, we relate the computation with the case of the previous \virgolette flat'' case of the Lie superalgebra $\mathpzc{susy} (\mathbb{R}^{1,2|2})$ considered above.

%\subsubsection{Cohomology of $\mathfrak{osp}(1|2)$}

\noindent The choice of a basis for $\mathfrak{osp}(1|2)$ using the relations \eqref{relosp} is reflected into the commutation relations. However, a neat and convenient choice is provided as follows: 
\bear
\mathcal{P}_0 = \frac{1}{2}\left ( 
\begin{array}{c|cc}
0 & 0 & 0 \\
\hline 
0 & 0 & 1 \\
0 & - 1 & 0  
\end{array}
\right ), \qquad 
\mathcal{P}_1 = \frac{1}{2} \left ( 
\begin{array}{c|cc}
0 & 0 & 0 \\
\hline 
0 & 1 & 0 \\
0 & 0 & -1  
\end{array}
\right ), \qquad
\mathcal{P}_2 = \frac{1}{2}\left ( 
\begin{array}{c|cc}
0 & 0 & 0 \\
\hline 
0 & 0 & 1 \\
0 & 1 & 0  
\end{array}
\right ), \quad 
\eear
\bear
\mathcal{Q}_1 = \left ( 
\begin{array}{c|cc}
0 & 1 & 1 \\
\hline 
1 & 0 & 0 \\
-1 & 0 & 0  
\end{array}
\right ), \qquad
\mathcal{Q}_2 = \left ( 
\begin{array}{c|cc}
0 & 1 & -1 \\
\hline 
-1 & 0 & 0 \\
-1 & 0 & 0  
\end{array}
\right ).
\eear
Making use of the previously introduced (real and symmetric) gamma matrices $\gamma^i_{\a \b}$ the commutation relations can be written in the following very convenient way
\begin{align}
& [\mathcal{P}_a, \mathcal{P}_b] = -  \epsilon_{abc}\mathcal{P}_c, \qquad \{ \mathcal{Q}_\alpha, \mathcal{Q}_\beta \} = -2 \gamma^a_{\alpha\beta} \mathcal{P}_a, \qquad [\mathcal{Q}_\alpha , \mathcal{P}_a] = - \gamma^{\; \; a}_{\a \; \b} \mathcal{Q}_\b 
%[X_1, \Psi_1] = - \frac{1}{2} \Psi_1, \quad & [X_1, \Psi_2] = \frac{1}{2}\Psi_2, \quad
%[X_2, \Psi_1] = \frac{1}{2} \Psi_2, \nonumber \\
%[X_2, \Psi_2] = - \frac{1}{2}\Psi_1, \quad  & [X_3, \Psi_1] = - \frac{1}{2} \Psi_2, \quad [X_3, \Psi_2] = - \frac{1}{2}\Psi_1, \nonumber \\
%\{\Psi_1, \Psi_1 \} = X_2 + X_3, \quad &  \{ \Psi_1, \Psi_2 \} = \frac{1}{2} X_1, \quad \{\Psi_2, \Psi_2 \} = X_2 - X_3 
\end{align}
where $\epsilon_{abc}$ is the Levi-Civita symbol and where we observe that the first commutation relation follows by the isomorphism $\mathfrak{sp} (2, \mathbb{R}) \cong \mathfrak{so}(2,1, \mathbb{R}) \cong \mathfrak{su} (1,1, \mathbb{C})$. \\
%\begin{eqnarray}
%\label{roB}
%\{Q_\a, Q_\b\} = \gamma^a_{\a\b} P_a\,, ~~~~~~
%[Q_\a, P_a] = \gamma_{a, \a}^{~\b} Q_\b\,, ~~~~
%[P_a, P_b] = \epsilon_{ab}^{~c} P_c\,. 
%\end{eqnarray}
We now introduce the Maurer-Cartan forms which are dual to the above generators of the Lie superalgebra $\mathfrak{osp} (1|2)$ up to parity. More precisely we introduce a basis of forms such that $C^1_{CE, \mathpzc{dif}} (\mathfrak{osp} (1|2)) = \Pi \mathfrak{osp}(1|2)^\ast = \mbox{Span}_{\mathbb{R}} \{ \psi^\alpha | \mathpzc{V}^a \}$ for $a = 0, 1, 2$ and $\alpha = 1,2$ with $\mathpzc{V}^a (\pi \mathcal{P}_b) = \delta^a_b$ and $\psi^\alpha (\pi \mathcal{Q}_\beta) = \delta^\alpha_\beta$. The above commutation relations lead to the following set of Maurer Cartan equations (up to a sign redefinition):
\begin{align}
d \mathpzc{V}^a =  \epsilon_{bc}^{\; \; \; a} \mathpzc{V}^b  \mathpzc{V}^c + \psi^\a \gamma^a_{\a \b} \psi^\b, \qquad d \psi^\alpha =  \mathpzc{V}^a \gamma_{a, \a \b}  \psi^\b 
\end{align}
The cohomology reads 
\begin{align}
&H^0_{CE, \mathpzc{dif}} (\mathfrak{osp}(1|2)) \cong \mathbb{R} \cdot 1, \nonumber \\
& H^1_{CE, \mathpzc{dif}} (\mathfrak{osp}(1|2)) \cong 0, \nonumber \\
& H^2_{CE, \mathpzc{dif}} (\mathfrak{osp}(1|2)) \cong 0  \nonumber \\
& H^3_{CE, \mathpzc{dif}} (\mathfrak{osp} (1|2)) \cong \mathbb{R} \cdot \bigg \{ \frac12  \mathpzc{V}^a (\psi \gamma_a \psi) - \frac16 \epsilon_{abc} \mathpzc{V}^a \mathpzc{V}^b \mathpzc{V}^c\ \bigg \}
\end{align}
and $H^{p>3}_{CE, \mathpzc{dif}} (\mathfrak{osp} (1|2)) \cong 0.$ 
%{\bf are the constants correct? Maybe 1/3? I am most probably wrong...! Carlo: I think that the correct constants are 1/2 and -1/6, or 1 and -1/3.} 
Notice that this result is confirmed by the theorem of Fuks, which states that the cohomology of $\mathfrak{osp} (1|2)$ is isomorphic to that of its bosonic subalgebra $\mathfrak{sp}(2, \mathbb{R})$, thus leading to the Poincar\'e polynomial 
\bear
\mathpzc{P}_{\mathfrak{osp} (1|2)}^{\mathpzc{dif}} [t] = \mathpzc{P}_{\mathfrak{sp} (2, \mathbb{R})} = 1 - t^3.
\eear
Notice, though, that with respect to the bosonic Lie algebra $\mathfrak{sp}(2, \mathbb{R})$ the representative of the $3$-cohomology of the Lie superalgebra $\mathfrak{osp}(1|2)$ is shifted in the fermionic directions as can be seen directly by the above expression. \\
Quite similarly, the integral Chevalley-Eilenberg cohomology reads
\begin{align}
&H^3_{CE, \mathpzc{int}} (\mathfrak{osp}(1|2)) \cong \mathbb{R} \cdot 
 \epsilon_{abc} \mathpzc{V}^a \mathpzc{V}^b \mathpzc{V}^c \epsilon_{\a \b} \delta (\psi^\a) \delta (\psi^\b), \nonumber \\
& H^2_{CE, \mathpzc{int}} (\mathfrak{osp}(1|2)) \cong 0, \nonumber \\ 
& H^1_{CE, \mathpzc{int}} (\mathfrak{osp}(1|2)) \cong 0,  \nonumber \\
& H^0_{CE, \mathpzc{int}} (\mathfrak{osp} (1|2)) \cong \mathbb{R} \cdot \bigg \{ \frac12  
\mathpzc{V}^a \mathpzc{V}^b (\iota_{\pi \mathcal{Q}_\a} \gamma_{[ab], \a \b} 
\iota_{\pi \mathcal{Q}_\b}) \epsilon_{\a \b} \delta (\psi^\a) \delta (\psi^\b) - \frac16 \epsilon_{\a \b} \delta (\psi^\a) \delta (\psi^\b)  \bigg \} \ .
\end{align}
%{\bf Sistemare costanti come in 4.115} 
It is worth to observe the relation between the \virgolette curved'' and \virgolette flat'' 3-dimensional case. Indeed, simply redefining the generators of the superalgebra $\mathfrak{osp}(1|2)$ by a constant parameter $\lambda$ as follows,
\bear
\mathcal{Q}^{\lambda}_\alpha \defeq \frac{1}{\sqrt{\lambda}} \mathcal{Q}_\alpha, \qquad \mathcal{P}_{a}^{\lambda} \defeq \frac{1}{\lambda} \mathcal{P}_a,
\eear
one finds that the new Maurer-Cartan equations for $\mathpzc{V}_{\lambda}^a$ and $\psi^\alpha_\lambda $ read
\bear
d \mathpzc{V}^a_{\lambda} = \lambda  \epsilon_{bc}^{\; \; \; a} \mathpzc{V}^b_{\lambda}  \mathpzc{V}^c_{\lambda} +  \psi^\a_{\lambda} \gamma^a_{\a \b} \psi^\b_{\lambda}, \qquad d \psi_\lambda^\alpha = \lambda  
\mathpzc{V}^a \gamma_{a, \a \b}. 
\eear
The limit $\lambda \rightarrow 0$ is called \emph{\.{I}n\"on\"u-Wigner contraction} and it is immediate to see that it gives back the Maurer-Cartan equations for the superalgebra $\mathpzc{susy} (\mathbb{R}^{1,2|2}):$ in this sense $\mathpzc{susy} (\mathbb{R}^{1,2|2})$ can be seen as the \virgolette flat'' limit of the orthosymplectic superalgebra $\mathfrak{osp}(1|2)$.

\subsection{Dimension 4: \virgolette Flat'' and \virgolette Curved'' Cases}

\subsubsection{Flat Case: Supertranslations of the $D=4$, $\mathcal{N}=1$ Superspace $\mathbb{R}^{1,3|4}$}

\noindent Let us now move to a 4 dimensional example. Here we consider the physically relevant superspace $\mathbb{R}^{1,3|4}$ based upon the 4-dimensional Minkowski space $\mathbb{R}^{1,3}$. This is the usual superspace 
for rigid supersymmetry models  $\mathcal{N}=1$ and therefore the first step toward supergravity models. 
Some of results of the present section have been also discussed in \cite{Brandt:1992ts}. 

Again, we describe this flat supermanifold via the coordinates $(x^a | \theta^\a)$ for $a=0,\ldots 3$ and $\a = 1, \ldots 4$. The supersymmetry generators read exactly as in equation \eqref{3dimgen}, but clearly now the gamma's are 4-dimensional Dirac matrices, instead of 2-dimensional. Accordingly, passing to the Maurer-Cartan forms and defining $\mathpzc{V}^a = dx^a + %\sum_{\alpha, \beta }
\theta^\a \gamma^a_{\a \b} d\theta^\beta$ and $\psi^\a = d\theta^\a $ one has that the generators of the 1-cochains of the Lie superalgebra satisfies the Maurer-Cartan equations
\bear
d\mathpzc{V}^a = \psi^\a \gamma^a_{\a \b} \psi^\beta, 
\qquad d\psi^\a = 0\,. 
\eear
Notice, by the way, that it is convenient switching the reducible Dirac representation $\psi \in (1/2, 0) \oplus(0, 1/2)$ to its irreducible components, the (left) Weyl spinors $\chi^\a \in (1/2,0)$ and (right) anti-Weyl spinors $\bar \lambda^{\dot \a} \in (0, 1/2)$ for $\a, \dot \a = 1, 2 $ so that $\psi = (\chi^\a , \bar \lambda^{\dot \a })$. The above Maurer-Cartan modifies to 
\bear
d\mathpzc{V}^{\a \dot \a} = \chi^\a \bar \lambda^{\dot \a}, \qquad  d\chi^\a = 0, \qquad d \bar \lambda^{\dot \alpha } = 0. 
\eear
Here we are using the spin structure to represent the odd 1-forms $\mathpzc{V}^a$ as bispinors: $\mathpzc{V}^{\alpha \dot{\alpha}} = \bar{\sigma}_{a}^{\alpha \dot{\alpha}} \mathpzc{V}^a $, where we have used the matrices $\bar{\sigma}$ of the $(0, 1/2)$ irreducible component. Instead of giving the cohomology classes, let us first look at the Poincar\'e polynomial. Assigning weights $1/2$ to the Maurer-Cartan forms $(\chi^\a, \bar \lambda^{\dot \a})$ and $1$ to the Maurer-Cartan form $\mathpzc{V}^i$ respectively, as already done early on, one considers
\bear
\hspace{-0.4cm}\mathpzc{P}_{\mathpzc{susy} (\mathbb{R}^{1,3|4})}^{\mathpzc{dif}} [\sqrt{t}] = \frac{(1-t)^4}{(1-\sqrt{t})^4} = \frac{(1-\sqrt{t})^4 (1+ \sqrt{t})^4}{(1-\sqrt{t})^4} = 
 1 + 4 \sqrt{t} + 6 t + 4 t \sqrt{t}  + t^2.
\eear
Here the numerator corresponds to product of the $\mathpzc{V}^{\a \dot \a}$'s - these are odd forms, thus they appear in the numerators - and the denominator corresponds to product of the $\chi^\a$'s and $\bar \lambda^{\dot \a}$'s - these are even forms, thus they appear in the numerator. Let us now see explicitly the first cohomology groups: 
\begin{align} 
& H^0_{CE, \mathpzc{dif}} (\mathpzc{susy}(\mathbb{R}^{1,3|{4}})) \cong \mathbb{R} \cdot 1, \nonumber \\
& H^1_{CE, \mathpzc{dif}} (\mathpzc{susy}(\mathbb{R}^{1,3|{4}})) \cong \mathbb{R} \cdot \{ \chi^\a, \; \bar\lambda^{\dot\a}\} \nonumber  \\
& H^2_{CE, \mathpzc{dif}} (\mathpzc{susy}(\mathbb{R}^{1,3|{4}})) \cong \mathbb{R} \cdot \{   \chi^\a \chi^\b, \; \bar\lambda^{\dot\a} \bar\lambda^{\dot\b}, \; \chi^\a\epsilon_{\a\b} V^{\b\dot\b},  \; \bar\lambda^{\dot\a}\epsilon_{\dot\a\dot\b} V^{\b\dot\b} \}, \nonumber \\
& H^3_{CE, \mathpzc{dif}} (\mathpzc{susy}(\mathbb{R}^{1,3|{4}})) \cong \mathbb{R} \cdot \{  
\chi^\a \chi^\b\chi^\gamma, \, 
\bar\lambda^{\dot\a} \bar\lambda^{\dot\b} \bar\lambda^{\dot\g}, \,
\chi^\gamma \chi^\a\epsilon_{\a\b} V^{\b\dot\b},  \,
\bar\lambda^{\dot\gamma}\bar\lambda^{\dot\a}\epsilon_{\dot\a\dot\b} V^{\b\dot\b}, \,
\chi^\a \bar\lambda^{\dot\a} \epsilon_{\a\b} \epsilon_{\dot\a\dot\b}  V^{\b\dot\b}  
\}, \nonumber \\ 
& H^4_{CE, \mathpzc{dif}} (\mathpzc{susy}(\mathbb{R}^{1,3|{4}})) \cong \mathbb{R} \cdot \{ \chi^\a \chi^\b\chi^\gamma \chi^\delta, \,
\bar\lambda^{\dot\a} \bar\lambda^{\dot\b} \bar\lambda^{\dot\g}\bar\lambda^{\dot\delta}, \dots \}
\end{align}
where the sum over repeated indices is understood, and the ellipses in the 4-cohomology group stays for other cohomology representatives which we have not written. Notice that the cohomology is again infinite dimensional, for example one has that 
\bear
\chi^{\a_1} \ldots \chi^{\a_p} \in H^p_{CE, \mathpzc{dif}} (\mathpzc{susy}(\mathbb{R}^{1,3|{4}}))
\eear
for any $p\geq 1$.
With reference to the assigned weights, one sees that the Poincar\'e series (in $\sqrt{t}$) is reconstructed as follows:  
\begin{align}
& \dim H^0_{CE, \mathpzc{dif}} (\mathpzc{susy}(\mathbb{R}^{1,3|{4}})) \rightsquigarrow 1, \nonumber \\
& \dim H^1_{CE, \mathpzc{dif}} (\mathpzc{susy}(\mathbb{R}^{1,3|{4}})) \rightsquigarrow 2\sqrt{t} + 2\sqrt{t}, \nonumber \\
& \dim H^2_{CE, \mathpzc{dif}} (\mathpzc{susy}(\mathbb{R}^{1,3|{4}})) \rightsquigarrow - 2t \sqrt{t} - 2t \sqrt{t} + 3 t + 3 t \nonumber \\
& \dim H^3_{CE, \mathpzc{dif}} (\mathpzc{susy}(\mathbb{R}^{1,3|{4}})) \rightsquigarrow 4t \sqrt{t} + 4t \sqrt{t} - 4 t^2  - 4 t^2 - t^2  \nonumber \\
& \dim H^4_{CE, \mathpzc{dif}} (\mathpzc{susy}(\mathbb{R}^{1,3|{4}})) \rightsquigarrow 5 {t}^2 + 5 {t}^2 + \ldots. 
%& \dim H^2_{CE, \mathpzc{dif}} (\mathpzc{susy}(\mathbb{R}^{1,3|{4}})) \rightsquigarrow 4t \sqrt{t} + 4t \sqrt{t} - 4 t^2  - 4 t^2 - t^2 \\
\end{align}
Summing up the above terms, this leads to $
 1 + 4 \sqrt{t} + 6 t + 4 t \sqrt{t}  + t^2 = \mathpzc{P}_{\mathpzc{susy} (\mathbb{R}^{1,3|4})}^{\mathpzc{dif}} $. Comparing explicit computations with the above Poincar\'e polynomial one indeed sees that the contributions for weights higher than 2 vanishes, or better, they sum up to zero, even if there is cohomology at any degree higher than 4. \\
 \noindent  However, the Maurer-Cartan equations allow to take different weights, namely distinguish between the left spinors and the right spinors, and associating to the $\chi$'s the weight $\sqrt t$ and to $\bar \lambda$'s the weight $\sqrt{\bar t}$, so that $\mathpzc{V}$ is associated with $\sqrt{t} \sqrt{\bar t}$. This choice leads indeed to the series:
 \bear
 \hspace{-0.5cm}\mathpzc{P}_{\mathpzc{susy} (\mathbb{R}^{1,3|4})}^{\mathpzc{dif}} ( \sqrt{t}, \sqrt{\bar{t}} ) = 1 + 2 (\sqrt{t} + \sqrt{\bar t}) + 3 (t + \bar t) + 4(t \sqrt{t} + t \sqrt{t}) - 2 (\bar t \sqrt{t} + t \sqrt{\bar t}) + \ldots \nonumber \\
 \eear
where each monomial of the series is in a one-to-one correspondence with a cohomology representative and the signs stand for even and odd parity. 

\subsection{Curved Case: Lie Superalgebra $\mathfrak{osp}({2|2})$} 

\noindent Before discussing coset superspaces - which appear more suitable to provide 
useful example of supergravity backgrounds - we consider a \virgolette curved'' 4 dimensional case, studying the cohomology of the Lie 
superalgebra $\mathfrak{osp}({2|2}) = C(2)$. The setting is given exactly as above and we refer to Appendix \ref{appB}. Before we start, though, it is useful to stress that the related Lie supergroup $OSp(2|2)$ cannot be given an interpretation from Minkowskian point of view, since it breaks the $SO(1,3)$-invariance to the subgroup $SO(2) \times Sp(2)$ - and indeed fermionic coordinates transforms under this subgroup. However, the example provides a useful comparison with the remarkable \virgolette flat superspace'' case above.  

%Again we assign the coordinates $\{x^a, x^0|\theta^\a_I\}$ where $a=1,2,3$ $\alpha=1,2$ and $I=1,2$. The fermionic coordinates $\theta^I_\a$ 
%carry a bifundamental representation of $SO(2)\times Sp(2)$. This specific group manifolds does not have 
%an interpretation from Minkowskian point of view, since it breaks $SO(1,3)$ invariance to the subgroup 
%$SO(2) \times Sp(3)$, however it is a useful comparison term with the \virgolette flat superspace'' case above.  

\noindent The Maurer-Cartan forms are $\mathpzc{V}^{a} = \gamma^a_{\a\b} \mathpzc{V}^{\a\b}$, $\mathpzc{V}^0$ and $\psi^\a_I$, having separated a \virgolette time'' direction. They satisfy the 
Maurer-Cartan equations 
\begin{eqnarray}
\label{MCciccioA}
d \mathpzc{V}^{\a\b} &=& (\mathpzc{V}\wedge \mathpzc{V})^{\a\b} + \psi^\a_I \eta^{IJ} \psi^\b_J\,, \nonumber \\
d \mathpzc{V}^{0} &=&- \epsilon_{\a\b} \psi^\a_I \epsilon^{IJ} \psi^\b_J\,, \nonumber \\
d \psi^\a_I &=& (V\wedge \psi)^\a_I + \epsilon_{I}^J \mathpzc{V}^0 \psi^\a_J\,. 
\end{eqnarray}
Notice that in the suitable \virgolette flat'' limit, one retrieves the 
flat model discussed in the previous section. According to Fuks, the cohomology should match with the one of the $\mathfrak{sp}({2}, \mathbb{R})$ subalgebra, and therefore we expect the Poincar\'e polynomial to be of the form 
\begin{eqnarray}
\label{MCciccioB}
\mathpzc{P}_{ \mathfrak{osp}({2|2})}^{\mathpzc{dif}}[t] = (1-t^3). 
\end{eqnarray} 
The cohomology generators are indeed found to read
\begin{eqnarray}
\label{MCciccioB}
H^{0}_{CE, \mathpzc{dif}}(\mathfrak{osp}({2|2})) &=& \mathbb{R} \cdot 1 \,, \nonumber \\
H^{3}_{CE, \mathpzc{dif}}(\mathfrak{osp}({2|2})) &=&  \mathbb{R} \cdot \left\{\psi^\a_I \eta^{IJ} \psi^\b_J  \mathpzc{V}_{\a\b} + 
\psi^\a_I \epsilon^{IJ} \psi^\b_J \epsilon_{\a\b} \mathpzc{V}^0 +  \mathpzc{V}\wedge  \mathpzc{V}\wedge  \mathpzc{V}\right\}
\end{eqnarray} 
On the other hand, cohomology classes in the integral form sector are explicitly given by 
\begin{eqnarray}
\label{MCciccioC}
H^{1}_{CE, \mathpzc{int}}(\mathfrak{osp}({2|2}) &=& \mathbb{R} \cdot \left\{ 
\iota_\a^I \eta_{IJ} \iota_\b^J \mathpzc{V}^0 ( \mathpzc{V} \wedge \mathpzc{V} )^{\a\b} \delta^4(\psi) + 
\iota_\a^I \epsilon_{IJ} \iota_\b^J \epsilon^{\a\b} (\mathpzc{V} \wedge \mathpzc{V} \wedge \mathpzc{V}) \delta^4(\psi) +  \mathpzc{V}^0 \delta^4(\psi) \right\} \nonumber \\
H^{4}_{CE, \mathpzc{int}}(\mathfrak{osp}({2|2}) &=& \mathbb{R} \cdot \left\{ 
 \mathpzc{V}^0 \mathpzc{V}\wedge  \mathpzc{V}\wedge  \mathpzc{V} \delta^4(\psi) \right\}. 
\end{eqnarray}
%It is easy to verify that the cohomology is isomorphic to cohomology of the superforms (\ref{MCciccioB}).  

\section{Coset Superspaces and \\ Equivariant Chevalley-Eilenberg Cohomology}

\noindent In this section we briefly introduce \emph{equivariant} Chevalley-Eilenberg cohomology, a crucial tool to study the cohomology of \emph{coset} or \emph{homogeneous superspaces} $\mathcal{G} / \mathcal{H}$ where $\mathcal{G}$ is a Lie supergroup and $\mathcal{H}$ is a Lie sub-supergroup of $\mathcal{G}$.

Very few examples of Lie supergroup, or \emph{group supermanifolds}, are indeed solutions of supergravity/string equations of motion, for example 
$AdS_3$ in the case of non-critical strings and few others. Nonetheless, the space of geometric backgrounds modelled on coset spaces is much richer, in particular the case of supersymmetric background built on \emph{coset supermanifolds}. 
In this context, the most important instance is that of a coset supermanifold realized by modding out a certain \emph{bosonic subgroup}: the 
infamous examples of $AdS_5 \times S^5$ and $AdS_4 \times \mathbb{CP}^3$ belong this category \cite{MT} \cite{Sorokin}.  Furthermore, a less explored instance it that obtained by modding out a true \emph{Lie sub-supergroup}. In any of each cases, it is interesting to compute their (equivariant) cohomology, as it can uncover insights in the physics related to the model.

\noindent Given a Lie supergroup $\mathcal{G}$ and a Lie sub-supergroup $\mathcal{H}$ of $\mathcal{G}$ we define the related Lie superalgebras by $\mathfrak{g}$ and $\mathfrak{h}$. Then, attached to the coset superspace $\mathcal{G}/ \mathcal{H}$ we will have, correspondingly, the quotient $\mathfrak{g} / \mathfrak{h}$, whose elements are equivalence classes $ g\, \mbox{mod} \, \mathfrak{h}$. As a vector superspace, there always exists a direct linear decomposition of $\mathfrak{g}$ such that 
\bear
\mathfrak{g} = \mathfrak{h} \oplus \mathfrak{C},
\eear
but the choice of $\mathfrak{C}$ is ambiguous and different compatibility conditions between this direct linear decomposition and the Lie algebra structures can be imposed. More in details, the coset superspace $\mathcal{G} / \mathcal{H}$ is said to be \emph{reductive} if there exists an $\mbox{ad}(\mathfrak{h})$-invariant choice of $\mathfrak{C}$, \emph{i.e.}\ 
\bear \label{reductive}
\mbox{ad} (\mathfrak{h}) \cdot \mathfrak{C} = [\mathfrak{h}, \mathfrak{C}] \subset \mathfrak{C}.
\eear
Further, imposing that $[\mathfrak{C}, \mathfrak{C}] \subset \mathfrak{h}$ we get that the coset $\mathcal{G} / \mathcal{H}$ is a \emph{symmetric superspace}, but in the following we will consider the more general relation 
\bear \label{gener}
[\mathfrak{C}, \mathfrak{C}] \subset \mathfrak{g}.
\eear 
As in the ordinary setting, left-translation in the coset superspace induces a map $(\ell_{[g^{-1}]})_\ast : \mathcal{T}_{[g]} \mathcal{G} / \mathcal{H} \rightarrow \mathcal{T}_{[e]} \mathcal{G} / \mathcal{H} \cong \mathfrak{g}/ \mathfrak{h}$ which can be seen as $\mathfrak{g} / \mathfrak{h}$-valued 1-forms, the so-called Maurer-Cartan forms. As above, we will always deal with \emph{matrix} Lie superalgebras. In this case the Maurer-Cartan is usually written starting from the coset superspace elements as $\omega_{MC}^g = [g^{-1} dg].$ Notice that, choosing another representative $gh $ for $h \in \mathcal{H}$ instead of $g$, we get
\bear
\omega^{gh}_{MC} = [\mbox{ad} (h) (g^{-1} dg)] = \mbox{ad} (h)\cdot \omega^g_{MC},
\eear
since $[h^{-1} dh] = 0 $ in the quotient $\mathfrak{g}/\mathfrak{h}$. Passing from the above coordinate-invariant formalism to a particular choice of coordinates, in line with the general philosophy of the paper of finding explicit expressions, we choose a certain direct linear decomposition of $\mathfrak{g}$ as above and, in turn, a basis $\{ h_i \}$ for $i = 1, \ldots, \dim \mathfrak{h}$ of generators for $\mathfrak{h}$ and a basis $\{ {k}_J \}$ for $J = 1, \ldots, \dim \mathfrak{C}$ of generators for $\mathfrak{C}$. Notice that the parametrization of the elements of the coset superspace $[g] \in \mathcal{G} /\mathcal{H}$ is far from being unique. The Maurer-Cartan form related to this decomposition and choice of basis can be computed as to get
\bear
\omega_{MC} = \mathpzc{V}^J k_J + \omega^i h_i,
\eear
where the $\mathpzc{V}^i$'s are the supervielbein forms and the $\omega^j$'s are interpreted as the connection forms associated with the action of the sub-superalgebra $\mathfrak{h}.$ The vielbein and connection forms satisfy the following Maurer-Cartan equations that can be read off the \eqref{reductive} and \eqref{gener}
\begin{align}
\label{EqCartan}
d \mathpzc{V}^I = f^I_{JK} \mathpzc{V}^J \wedge \mathpzc{V}^K + f^I_{i J} \, \omega^i\wedge \mathpzc{V}^J, \nonumber \\
d\omega^i = f^i_{jk} \omega^j \wedge \omega^k + f^i_{IJ} \mathpzc{V}^I \wedge \mathpzc{V}^K.
\end{align}
The second one can be re-written as
\bear
\mathpzc{R}^i \defeq  d\omega^i - f^i_{jk} \omega^j \wedge \omega^k = f^i_{IJ} \mathpzc{V}^I \wedge \mathpzc{V}^K. 
\eear
Here the structure constants are written with respect to the above decomposition of the Lie superalgebra $\mathfrak{g} = \mathfrak{h} \oplus \mathfrak{C}$ and $\mathpzc{R}^i$ is referred to as the \virgolette curvature'' of the gauge connection $\omega^i$ related to the sub-algebra $\mathfrak{h}$. The form of the first Maurer-Cartan equation in \eqref{EqCartan} in turn makes convenient to introduce a \emph{covariant differential} defined as
\bear\label{covaD}
\mathcal{D} \mathpzc{V}^I \defeq d \mathpzc{V}^I - f^I_{i J} \, \omega^i\wedge \mathpzc{V}^J.
\eear
Notice that this differential is \emph{not} nilpotent, indeed one has
\bear \label{lieder}
\mathcal{D}^2 \mathpzc{V}^I = - \mathpzc{R}^i f_{iJ}^{I} \mathpzc{V}^J
\eear
using Jacobi identity. This can be re-written as $\mathcal{D}^2 \mathpzc{V}^I = - {\mathcal L}_\mathpzc{R} V^I$, where we have denoted $\mathcal{L}_{\mathpzc{R}}$ the action of the Lie derivative on the vielbeins $\mathpzc{V}^I$ along the (vertical) vector $\mathpzc{R}^i h_i$. \\ The above expression makes clear that in order to have a well-defined differential cochain complex for coset superspaces, we need to impose further conditions on the forms to take into account. Namely, we need the Maurer-Cartan forms, call them $\Omega$'s, to be \emph{basic}, this means that we require  
\bear
\mathcal{L}_{H} \Omega = 0, \qquad \iota_{H} \Omega = 0,  
\eear
for any vector $H$ coming from the sub-algebra $\mathfrak{h}$. Roughly speaking, one can visualize these requirements thinking about a principal $\mathcal{H}$-bundle $\pi : \mathcal{P} \rightarrow \mathcal{G} / \mathcal{H}$: in this respect a \emph{basic form} $\Omega$ is a form defined on the principal bundle $\Omega = \pi^\ast (\mathpzc{V})$ such that it has \emph{no vertical components} (it is {horizontal}) and \emph{no vertical variation} (it stays horizontal), \emph{i.e.}\ basic forms are in the intersection $\ker (\iota_{H} )\cap \ker (\mathcal{L}_H)$. Calling $C^p_{\mathpzc{Basic}} (\mathfrak{g}/\mathfrak{h})$ the (vector) superspace of the basic $p$-forms, we accordingly define the equivariant (Chevalley-Eilenberg) cohomology to be the cohomology of the basic forms with respect to the differential $\mathcal{D}$ introduced above. 
\bear
{H}^{p}_{\mathpzc{EQ}} ({\mathfrak{g} / \mathfrak{h}})  \defeq \frac{\{ \Omega \in C^p_{\mathpzc{Basic}} (\mathfrak{g} / \mathfrak{h}) : \mathcal{D} \Omega = 0 \}}{\{ \Omega \in C^{p}_{\mathpzc{Basic}} (\mathfrak{g}/ \mathfrak{h}) : \exists \eta \in C^{p-1}_{\mathpzc{Basic}} (\mathfrak{g} / \mathfrak{h}) \; \Omega = \mathcal{D} \eta \}} .
\eear

\subsection{Methods for Computations: Poincar\'e Polynomial Revised}

\noindent In absence of encompassing \virgolette structure theorems'', different methods are possible in order to compute cohomology of coset superspaces. Our strategy will be to supplement brute force computations with the indications coming from the Poincar\'e polynomial of coset superspaces. This will tell, for example, when a cohomology space is expected to be infinite dimensional, as we shall see. \\
Following \cite{greub}, if $\mathfrak{g}$ is a Lie superalgebra with Poincar\'e series given by $\mathpzc{P}_\mathfrak{g} (t) = \sum_i b_i^\mathfrak{g} t^i$, where the $b_i^\mathfrak{g}$'s are the Betti numbers of the Lie superalgebra $\mathfrak{g}$, \emph{i.e.} $b^\mathfrak{g}_i \defeq \dim H^i_{CE} (\mathfrak{g})$ and $\mathfrak{h}$ is a Lie sub-superalgebra of $\mathfrak{g}$, of the same rank (Cartan Pairs, see \cite{greub}) having Poincar\'e series given by $\mathpzc{P}_\mathfrak{h} (t) = \sum_j b_i^\mathfrak{h} t^j$, then the Poincar\'e series for the coset will be given by the following formula
\bear \label{cosetp}
\mathpzc{P}_{\mathfrak{g}/\mathfrak{h}}(t) = \frac{\sum_{i} b^\mathfrak{g}_i t^{i+1}}{\sum_{j} b^\mathfrak{h}_j t^{j+1}} = 
\frac{\prod_{l} (1-t^{c^\mathfrak{g}_l+1})}{\prod_{m} (1-t^{c^\mathfrak{h}_m+1})}  
\eear
where $c^{\mathfrak{g}}_l$ and $c^{\mathfrak{h}}_m$ are the usual exponents in the factorised form of the polynomial. This product formula is very helpful since it provides some informations regarding the 
different cohomology classes. 

A remark is in order: for ordinary coset spaces arising from purely bosonic finite dimensional Lie algebras $\mathfrak{h} \subset \mathfrak{g}$  the above formula \eqref{cosetp} always yields a polynomial: $\mathpzc{P}_{\mathfrak{h}} (t)$ divides $\mathpzc{P}_\mathfrak{g} (t)$ (which is a polynomial), or equivalently the $c^{\mathfrak{h}}_m$ are a subset of the $c^{\mathfrak{g}}_l$. For example, a $2n$-dimensional sphere $S^{2n}$ can be seen as the coset manifold given by the quotient of Lie groups $SO(2n+1) / SO(2n)$. In the simplest case $n=1$ we obtain a 2-sphere $S^2$: passing to the algebras we have the coset $\mathfrak{so}(3)/\mathfrak{so}(2)$ which is isomorphic to the coset $\mathfrak{su}(2)/ \mathfrak{u}(1)$. The vielbeins $\{ \mathpzc{V}^0, \mathpzc{V}^{+}, \mathpzc{V}^{-} \}$ for $\mathfrak{so}(3)\cong \mathfrak{su}(2)$ satisfies the following Maurer-Cartan equations
\begin{eqnarray}
\label{fucD}
d \mathpzc{V}^0 = \mathpzc{V}^+ \wedge \mathpzc{V}^-\,, ~~~~~ d \mathpzc{V}^+ = i \mathpzc{V}^0 \wedge \mathpzc{V}^+\,, ~~~~ d \mathpzc{V}^- = - i \mathpzc{V}^0 \wedge V^+.
\end{eqnarray}
These can be rewritten as
\begin{eqnarray}
\label{fucD}
\mathpzc{R} = d \mathpzc{V}^0 = \mathpzc{V}^+ \wedge \mathpzc{V}^-, ~~~~~ \mathcal{D} \mathpzc{V}^\pm = 0\,, ~~~~ 
\end{eqnarray}
where $\mathcal{D}$ is as above and the Bianchi identities imply that $\mathcal{D} \mathpzc{R} =0$ and $\mathcal{D}^2 V^\pm = \pm i \mathpzc{R} \wedge \mathpzc{V}^\pm =0$. Notice that the nilpotent of $\mathcal{D}$ holds true since since $\mathpzc{R} = \mathpzc{V}^+ \wedge \mathpzc{V}^-$ and 
$\mathpzc{V}^\pm \wedge \mathpzc{V}^\pm =0$. 
In addition, we notice that ${\mathcal L}_\mathpzc{T} \mathpzc{V}^\pm = i \mathpzc{V}^\pm$ where $\mathpzc{T}$ is the generator of $\mathfrak{u}(1)$ subalgebra. This implies that the only basic forms are given by $\{1, \mathpzc{V}^+ \wedge \mathpzc{V}^-\}$, indeed notice also that $\iota_\mathpzc{T} \mathpzc{V}^0 =1 \neq 0$, thus $\mathpzc{V}^0$ is not a basic form. The basic forms 
$\{1, \mathpzc{V}^+ \wedge \mathpzc{V}^-\}$ are closed and not exact. The first statement is obvious. For the latter we observe 
that $\mathpzc{V}^+ \wedge \mathpzc{V}^- = \mathcal{D} \mathpzc{V}^0$, with $\mathcal{D}$ given in \eqref{covaD} (notice that $\mathcal{D}\mathpzc{V}^0 = d \mathpzc{V}^0$) but since $\mathpzc{V}^0$ is not basic. It follows that $\mathpzc{V}^+ \wedge \mathpzc{V}^-$ defines indeed an equivariant cohomology class. \\
The Poincar\'e polynomial for $\mathfrak{so}(3)$ is given by $\mathpzc{P}_{\mathfrak{so}(3)} (t) = \mathpzc{P}_{\mathfrak{su}(2)} (t)= 1 - t^3$, while the subalgebra $\mathfrak{u}(1) \cong \mathfrak{so}(2)$ has Poincar\'e polynomial given by $\mathpzc{P}_{\mathfrak{u}(1)} (t) = 1-t$, so that according to the above formula the coset has Poincar\'e polynomial given by  %where $-t$ corresponds to the volume form on the circle. 
%To compute the Poincar\'e polynomial of the sphere $S^2$, we use the above-mentioned formula 
\begin{eqnarray}
\label{fucE} 
\mathpzc{P}_{\mathfrak{so}(3) / \mathfrak{so}(2)} [t] = \frac{1-t^4}{1-t^2} = 1 + t^2,
\end{eqnarray}
thus matching the above calculation. %Getting back to $S^2$ seen as the coset manifold by the namely the constant and the \virgolette volume'' form on the sphere $S^2$. 
In the next section we generalize this easy example to the case of the supersphere.

\subsection{Lower Dimensional Cosets of $\mathfrak{osp}(1|2)$ and $\mathfrak{u}(1|1)$}

\noindent Let us now consider the Lie superalgebra $\mathfrak{osp}(1|2)$ introduced above. In agreement with an early result by Fuks, we have seen that $H^\bullet_{CE} (\mathfrak{osp}(1|2)) \cong H^\bullet_{CE} (\mathfrak{sp} (2, \mathbb{R}))$ and in particular, its Poincar\'e polynomial reads $\mathpzc{P}_{\mathfrak{osp} (1|2)} (t) = 1-t^3$ with the 3-cohomology class generated by $\omega^{(3)} = \psi \gamma_a \psi \mathpzc{V}^a + \frac13 \epsilon_{abc} \mathpzc{V}^a \mathpzc{V}^b \mathpzc{V}^c$, where the vielbeins $\psi$'s and $\mathpzc{V}$'s have been introduced above together with the gamma matrices $\gamma^i_{\a \b}.$ \\
Looking at the Lie supergroup $OSp(1|2)$ related to $\mathfrak{osp}(1|2)$ it is natural to consider two coset manifolds. The first one is the coset $OSp(1|2) / SO(1,1)$, which is known as the \emph{supersphere} $S^{2|2}$. The second one is a purely fermionic superspace, actually a \virgolette fat point'', given by the coset $OSp(1|2) / Sp(2, \mathbb{R})$, which is a $0|2$-dimensional superspace.\\
Let us start from the supersphere: the Lie algebra coset Poincar\'e polynomial reads
\bear
\mathpzc{P}_{\mathfrak{osp}(1|2)/ \mathfrak{so}(1,1)} [t] = \frac{1-t^4}{1-t^2} = 1+t^2,
\eear
upon using the so-called \emph{Weyl trick}, in order to identify the Chevalley-Eilenberg cohomology of $\mathfrak{so}(1,1)$ with that of $\mathfrak{so}(2) \cong \mathfrak{u}(1).$ This suggests that besides the constants there is a single cohomology class at degree two. Explicitely, introducing the Maurer-Cartan vielbeins $\{ \mathpzc{V}^0, \mathpzc{V}^\ddagger, \mathpzc{V}^=| \psi^\pm \}$ - notice that the the $\mathpzc{V}$'s are odd forms and the $\psi$'s are even forms - one gets the following Maurer-Cartan equations:
\begin{align}
&\mathcal{D} \mathpzc{V}^0 = \mathpzc{R} = i \mathpzc{V}^\ddagger \wedge \mathpzc{V}^= + \psi^+ \wedge \psi^-\,, ~~~~
\mathcal{D} \mathpzc{V}^\ddagger = i \psi^+\wedge \psi^+\,, ~~~~
\mathcal{D} V^= = - i \psi^-\wedge \psi^-\,, ~~~~\nonumber \\
&\qquad \qquad \qquad \qquad \mathcal{D} \psi^+ = \mathpzc{V}^\ddagger \wedge \psi^- \,, ~~~~
\mathcal{D} \psi^- = \mathpzc{V}^= \wedge \psi^+ \,. ~~~~
\end{align}
The infinitesimal action of the subgroup is given by
\bear
\label{fucH}
{\mathcal L}_\mathpzc{T} V^\ddagger =  2 i V^\ddagger\,, ~~~~
{\mathcal L}_\mathpzc{T} V^= =  -2 i V^=\,, ~~~~
{\mathcal L}_\mathpzc{T} \psi^\pm = \pm i \psi^\pm. 
\eear
Note that the 2-form
\bear
\mathpzc{R} = i \mathpzc{V}^\ddagger \wedge \mathpzc{V}^= + \psi^+ \wedge \psi^-\,
\eear
is (real) basic and closed. It is not exact because $\mathpzc{R} = \mathcal{D} \mathpzc{V}^0$, but $\mathpzc{V}^0$ is not a basic form. Notice in particular that the $3$-cohomology class of $\mathfrak{osp} (1|2)$ is no longer a cohomology class of the coset; it is invariant under the action of the subgroup, but it is not basic. We have
\bear
H^p_\mathpzc{EQ}(\mathfrak{osp}(1|2) / \mathfrak{osp}(1|1)) = \left \{
\begin{array}{lll}
\mathbb{R} & \quad & p =0, 2\\
0 & \quad &  \mbox{else}.
\end{array}
\right.
\eear
In the case of the fermionic coset the Poincar\'e polynomial reads
\bear
\mathpzc{P}_{\mathfrak{osp}(1|2)/ \mathfrak{sp}(2)} (t) = \frac{1-t^4}{1-t^4} = 1.
\eear
We expect therefore only the constants be in the cohomology, which is indeed the case since now $\mathpzc{R}$ is \emph{not} basic as now the forms $\mathpzc{V}^{\ddagger}$ and $\mathpzc{V}^{=}$ are not vielbeins, but connections instead, coming from the subalgebra $\mathfrak{sp}(2)$:
\bear
H^p_\mathpzc{EQ}(\mathfrak{osp}(1|2) / \mathfrak{sp}(2, \mathbb{R})) = \left \{
\begin{array}{lll}
\mathbb{R} & \quad & p =0\\
0 & \quad &  \mbox{else}.
\end{array}
\right.
\eear

\noindent We now consider the case of $\mathfrak{u}(1|1)$, whose Chevalley-Eilenberg cohomology has been discussed above. Namely, we consider the coset $\mathfrak{u}(1|1) / \mathfrak{u}(1)$ of dimension $1|2$ and $\mathfrak{u}(1|1) / \mathfrak{u}(1)\oplus \mathfrak{u}(1)$ of dimension $0|2$.\\
Let us start from the first coset space. A subtle point is that we have to choose how to embed the abelian factor $\mathfrak{u}(1)$ inside $\mathfrak{u}(1|1)$: indeed the automorphism of $\mathfrak{u}(1|1)_{0} = \mathfrak{u}(1) \oplus \mathfrak{u}(1)$ that exchange the factors does not lift to $\mathfrak{u}(1|1)$ (see, e.g., \cite{Frappat:1996pb}). With reference to the previous section, we can embed $\mathfrak{u} (1)$ in such a way that its Maurer-Cartan form (the connection, in view of the equivariant cohomology) is associated either to $U$ or to $W$.
In the case it is associated to $U$, then the cohomology trivializes as can be readily observed from the Maurer-Cartan equations: the only non-zero equivariant cohomology group is the zeroth-cohomology group:
\bear
H^p_\mathpzc{EQ}(\mathfrak{u}(1|1) / \mathfrak{u}_U(1)) = \left \{
\begin{array}{lll}
\mathbb{R} & \quad & p =0\\
0 & \quad &  \mbox{else},
\end{array}
\right.
\eear
having called $\mathfrak{u}(1|1) / \mathfrak{u}_U(1)$ the related coset.\\
Things changes drastically in case $\mathfrak{u}(1)$ is embedded in a way such that its Maurer-Cartan forms correspond with $W$. In this case $U$ is the generator of a cohomology class, indeed it is closed and basic. Moreover, also the bilinears $(\psi^1 \psi^2)^p$ for any $p$ are in the equivariant cohomology: indeed they can be seen to be exact with respect to a non-basic term. The cohomology is therefore infinite-dimensional and generated by the elements $\{1, U \} \otimes \{ (\psi^1 \psi^2)^p\}$ for any $p\geq 0.$\\ 
\bear
H^p_\mathpzc{EQ}(\mathfrak{u}(1|1) / \mathfrak{u}_W(1)) = \left \{
\begin{array}{lll}
\mathbb{R} & \quad & p \mbox{ even}\\
\Pi \mathbb{R} & \quad &  p \mbox{ odd},
\end{array}
\right.
\eear
having called $\mathfrak{u}(1|1) / \mathfrak{u}_W(1)$ the related coset.\\
%The corresponding Poincar\'e polynomial reads 
%\bear
%\mathpzc{P}_{\mathfrak{u}(1|1)/ \mathfrak{u}(1), W} (t) = \frac{1-t}{1-t^2} = \frac{1}{1-t}.
%\eear
Finally, considering the coset $\mathfrak{u}(1|1) / \mathfrak{u}(1) \oplus \mathfrak{u}(1)$ we have that in this case both $U$ and $W$ correspond to Maurer-Cartan forms for the subalgebra. From the Maurer-Cartan equations it follows that the cohomology is generated by the representatives given by the fermionic bilinears $\{ (\psi^1 \psi^2)^p \},$ for any $p\geq 0$ so that the equivariant cohomology is non-zero in any even degree. 
\bear
H^p_\mathpzc{EQ}(\mathfrak{u}(1|1) / \mathfrak{u}{(1)} \otimes \mathfrak{u}{(1)}) = \left \{
\begin{array}{lll}
\mathbb{R} & \quad & p \mbox{ even}\\
0 & \quad &  p \mbox{ odd},
\end{array}
\right.
\eear
The corresponding Poincar\'e series read
\begin{equation}\label{U1ECD}
	\mathpzc{P}_{\mathfrak{u}(1|1) / \mathfrak{u}_W(1)}(t) = \frac{1-t}{1-t^2} = \frac{1}{1+t} 
	= 1 - t + t^2 - t^3\dots. 
	\ .
\end{equation}
and
\begin{equation}\label{U1ECF}
	\mathpzc{P}_{\mathfrak{u}(1|1) / \mathfrak{u}{(1)} \otimes \mathfrak{u}{(1)}}(t) = \frac{1-t^2}{\left( 1 - t^2 \right)^2} = \frac{1}{1- t^2} =
	1 + t^2 + t^4 + \dots
	\,. 
\end{equation}
which match the computations above.

\subsection{Higher Dimensional Cosets of 
$\mathfrak{osp}(2|2)$, $\mathfrak{osp}(3|2)$ and $\mathfrak{osp}(4|2)$} 

\noindent We now consider higher-dimensional cosets of $\mathfrak{osp}(n|2)$ for $n=2, 3, 4$. We start with some general considerations, and then we specialize to the single cases together with their cosets.\\ On a general ground, the Maurer-Cartan equations for $\mathfrak{osp}(n|2)$
\begin{align}
\label{ffB}
& d \mathpzc{V}_{(\a\b)} = \psi^I_\a \psi^J_\b \eta_{IJ} + (\mathpzc{V} \wedge \mathpzc{V})_{(\a\b)}\,, \nonumber \\
& d \mathpzc{T}^{[IJ]} = - \psi^I_\a \psi^J_\b \Omega^{\a\b} + (\mathpzc{T}\wedge \mathpzc{T})^{[IJ]}\,, \nonumber \\
& d \psi^I_\a = \mathpzc{V}_{\a\b} \Omega^{\b\gamma} \psi_\gamma^I + \mathpzc{T}^{[IJ]} \eta_{JK} \psi^K_\a
\end{align}
where the Maurer-Cartan forms are given by $\{ \psi^I_\a | \mathpzc{V}^{(\a \b)}, \mathpzc{T}^{[IJ]} \}$ for $\a, \b =1,2$ and $I,J,K, \ldots =1, \dots, n$. We have $ (\mathpzc{V} \wedge \mathpzc{V})_{(\a\b)} =  \mathpzc{V}_{(\a\gamma)} \Omega^{\gamma\delta} \mathpzc{V}_{(\delta\b)}$ and 
$(\mathpzc{T}\wedge \mathpzc{T})^{[IJ]} = \mathpzc{T}^{[IK]} \eta_{KL} \mathpzc{T}^{[LJ]}$, where $\Omega^{\a\b}$ is the $2$-symplectic invariant tensor (from $\mathfrak{sp}(2)$) and 
$\eta_{IJ}$ is the Euclidean rotation-invariant metric (from $\mathfrak{so}(n)$). \\
For any $n$ the 3-form 
\bear\label{3form}
\hspace{-0.5cm}\omega^{(3)} = \psi^I_\a \psi^J_\b \eta_{IJ} \mathpzc{V}^{(\a\b)} + \psi^I_\a \psi^J_{\b} \Omega^{\a\b} \mathpzc{T}_{IJ} + 
(\mathpzc{V}\wedge \mathpzc{V}\wedge \mathpzc{V})_{\a\b} \Omega^{\a\b} + (\mathpzc{T}\wedge \mathpzc{T}\wedge \mathpzc{T})_{IJ} \eta^{IJ} 
\eear
gives an invariant which is a representative of the 3-cohomology group $H^3_{CE} (\mathfrak{osp}(n|2))$, shared by all of the $\mathfrak{osp}(n|2)$. This is the unique cohomology class up to $n=3$ (besides the constants in the 0-cohomology), indeed we have that 
\bear
\mathpzc{P}_{\mathfrak{osp} (2|2)} [t] = \mathpzc{P}_{\mathfrak{osp} (3|2)} [t]= \mathpzc{P}_{\mathfrak{osp} (1|2)} [t]  = \mathpzc{P}_{\mathfrak{sp} (2, \mathbb{R})} [t] = 1- t^3.
\eear 
Things start changing in the case of $\mathfrak{osp}(4|2)$, indeed in this case one has  that 
\bear
\mathpzc{P}_{\mathfrak{osp} (4|2)} [t] = \mathpzc{P}_{\mathfrak{so} (4) } [t] = (1-t^3)^2,
\eear
where we recall that $D_2 \cong A_1 \otimes A_1$ for the complexified algebras and the Poincar\'e polynomial for $A_1$ is indeed $1-t^3$. We therefore expect a further 3-form in the cohomology $H^3_{CE}(\mathfrak{osp}(4|2)).$ This is indeed the case and the extra cohomology representative is given by
\begin{eqnarray}
\label{ffD}
\widetilde{\omega}^{(3)} = \epsilon_{IJKL} \psi^I_\a \psi^J_\b \Omega^{\a\b} \mathpzc{T}^{KL} + \epsilon_{IJK[M} \eta_{N]L} \mathpzc{T}^{IJ} \mathpzc{T}^{KL} \mathpzc{T}^{MN}.
\end{eqnarray}
Cohomology classes for higher dimensional $\mathfrak{osp}(n|2)$ for $n>4$ can be constructed in similar way.

\subsubsection{$\mathfrak{osp}(2|2)$}

\noindent  Let us now get back to the specific case of the Lie superalgebra $\mathfrak{osp}(2|2)$. This is a Lie superalgebra of dimension $4|4$, whose reduced algebra is given by $\mathfrak{osp}(2|2)_0 = \mathfrak{so}(2) \oplus \mathfrak{sp}(2, \mathbb{R})$. We consider its cosets $\mathfrak{osp}(2|2)/ \mathfrak{so}(1,1)$ and $\mathfrak{osp}(2|2)/ \mathfrak{so}(2)\oplus \mathfrak{so}(1,1)$, respectively of dimension $3|4$ and $2|4$. While the Poincar\'e series for the first coset can not be guessed by \eqref{cosetp} (because the two superalgebras have different rank), it can be immediately written for the double coset:
\begin{align}
& \mathpzc{P}_{\mathfrak{osp}(2|2)/\left( \mathfrak{so}(2)\oplus \mathfrak{so}(1,1) \right)} (t) = \frac{(1-t^4)}{(1-t^2)^2}= \frac{1+ t^2}{1-t^2}.
\end{align}
Let us calculate explicitly the cohomology of the two cosets: by looking at the Maurer-Cartan equations one finds
\begin{align}
\label{fucL}
&d \mathpzc{V}^{(\a\b)} = (\mathpzc{V} \wedge \mathpzc{V})^{(\a\b)} + \psi^\a_i \wedge \psi^\b_j \eta^{ij} , \nonumber \\
&d \mathpzc{W} = \psi^\a_i \wedge \psi^\b_j \epsilon^{ij} \epsilon_{\a\b}, \nonumber \\
& d \psi^\a_i = \mathpzc{V}^{(\a\gamma)} \epsilon_{\gamma\beta}\wedge \psi^\b_i - \mathpzc{W} \epsilon_{ij} \wedge \psi^{\a i}\,.  
\end{align}
for $\alpha = 1,2$ and $i = 1, 2$, where $\eta^{ij}$ is the Minkowski metric.\\
In the case of the first coset $\mathfrak{osp}(2|2) / \mathfrak{so}(1,1)$, there are two ways to embed $\mathfrak{so}(1,1)$ in $\mathfrak{osp}(2|2)$: we can embed it in the $\mathfrak{sp} (2)$ part or in the $\mathfrak{so}(2)$ part (after a suitable signature redefinition via unitary trick, i.e. we can identify $ \mathfrak{so} (1,1) $ and $\mathfrak{so} (2) $). In the first case, one finds the cohomology class
\bear\label{osp(2|2)V0}
\mathpzc{R} = (\gamma^0)_{\a\b} ((\mathpzc{V} \wedge \mathpzc{V})^{(\a\b)} + \psi^\a_i \wedge \psi^\b_j \eta^{ij}) \ ,
\eear 
where $\gamma^0$ is the $0$-th Dirac gamma matrix: it is easy to check that this is indeed a basic closed and not exact form. To do this it is convenient to decompose the vielbeins as $\mathpzc{V}^{(\alpha \beta)} = \gamma_{a (\alpha \beta)} V^a, a \in \lbrace 0,\pm \rbrace$ (as in the previous section for $\mathfrak{osp} (1|2)$), then we are doing the quotient w.r.t. $\mathpzc{V}^0$. Hence \eqref{osp(2|2)V0} represent a form which is closed by construction, basic since it does not depend on $\mathpzc{V}^0$ and non-exact, being exact w.r.t. a non-basic object. In the second case, we have that the $\mathfrak{so}$ algebra is embedded in the $\mathfrak{so}$ subalgebra of $\mathfrak{osp}$, hence in this case we are doing the quotient w.r.t. $\mathpzc{W}$. In this case we immediately see from the MC equations \eqref{fucL} that the bilinear
\begin{eqnarray}
\label{fucM}
(\psi \cdot \psi) = \psi^\a_i \wedge \psi^\b_j \epsilon^{ij} \epsilon_{\a\b} = \mathcal{D} \mathpzc{W} \ ,
\end{eqnarray}
together with its powers, is a basic, closed, non-exact form.

\noindent On the other hand, we can study the second coset $\mathfrak{osp} (2|2) / \left( \mathfrak{so}(1,1) \oplus \mathfrak{so}(2)\right)$. In this case, either $\mathpzc{R}$ or the bilinear (and all its powers) $(\psi \cdot \psi)$ are part of the equivariant cohomology, making the cohomology infinite dimensional, generated by $\{1, \mathpzc{R} \} \otimes \{ (\psi \cdot \psi )^p\}$ for any $p$. The cohomologies that we have studied explicitly then read
\begin{align}
& H^p_\mathpzc{EQ}(\mathfrak{osp}(2|2) / \mathfrak{so}(1,1))_{\mathpzc{V}^0} = \left \{
\begin{array}{lll}
\mathbb{R} & \quad & p =0,  2\\
0 & \quad &  \mbox{else},
\end{array}
\right. \\
& H^p_\mathpzc{EQ}(\mathfrak{osp}(2|2) / \mathfrak{so}(1,1))_{\mathpzc{W}} = \left \{
\begin{array}{lll}
\mathbb{R} & \quad & p \mbox{ even}\\
0 & \quad &  \mbox{else},
\end{array}
\right. \\
& H^p_\mathpzc{EQ}(\mathfrak{osp}(2|2) / \left( \mathfrak{so}(1,1) \oplus \mathfrak{so}(2) \right)) = \left \{
\begin{array}{lll}
\mathbb{R} & \quad &  p = 0, \\
\mathbb{R}^2 & \quad & p = 2, 4, \ldots\\
0 & \quad &  \mbox{else}.
\end{array}
\right.
\end{align}

\subsubsection{$\mathfrak{osp}(3|2)$}

\noindent Let us look at the case of the cosets of $\mathfrak{osp}(3|2)$. Cosets by $\mathfrak{so}(2)$ or $\mathfrak{so}(1,1)$ works in pretty the same way as the above case of $\mathfrak{osp}(2|2)$. On the other hand, it is interesting to deal with the case $\mathfrak{osp}(3|2) / \mathfrak{so}(2) \oplus \mathfrak{so}(1,1)$. First, observe that the subgroup and the supergroup have the same rank, so by \eqref{cosetp} the Poincar\'e series reads
\bear
\mathpzc{P}_{\mathfrak{osp}(3|2) / \mathfrak{so}(2) \oplus \mathfrak{so}(1,1)} (t) = \frac{1+t^2}{1-t^2},
\eear
which is the same as in the case of the coset $\mathfrak{osp}(2|2) / \mathfrak{so}(2) \oplus \mathfrak{so}(1,1)$. However, in this case we run into a problem. Indeed two equivariant cohomology classes can be singled out:
\begin{align}
\label{codC}
& \mathcal{D} \mathpzc{V}_{0} = \psi^I_\a \psi^J_\b \eta_{IJ} \gamma_0^{\a\b} + \mathpzc{V}_+ \wedge \mathpzc{V}_-, \nonumber \\
& \mathcal{D} \mathpzc{T}^0 = - \psi^I_\a \psi^J_\b \Omega^{\a\b} s_{IJ } + \mathpzc{T}_+\wedge \mathpzc{T}_-, 
\end{align}
where $s_{IJ} = - s_{JI}$, which select ``direction" $\mathfrak{so}(3)$ in  denoted as $\mathpzc{T}_0$ - in pretty much the same way as the $\gamma^0$ allows to select a \virgolette direction'' in the Lie algebra $\mathfrak{sp}(2, \mathbb{R}).$ Notice that the above elements are not exact as $\mathpzc{V}^0$ and $\mathpzc{T}_0$ are not basic as they being the generators of the subgroup. This result seems contradicting the above Poincar\'e series computation though. Actually, to take into account the two independent cohomology classes above one needs to have a term of the kind $(1+t^2)^2$ in the numerator: we multiply and divide the above series by $1+t^2$ so that we get:
\bear
\mathpzc{P}_{\mathfrak{osp}(3|2) / \mathfrak{so}(2) \oplus \mathfrak{so}(1,1)} (t) = \frac{(1+t^2)^2}{1-t^4}. 
\eear
This suggests that the cohomology must be infinite, depending on a $4$-form which can indeed be found to be
\bear
\mathpzc{X}^{(4)} = \psi^I_{\a} \gamma_a^{\a\b} \psi^J_\b \eta^{ab} \epsilon_{R IJ} \eta^{RS} \epsilon_{S KL}\psi^K_{\a'} \gamma_b^{\a'\b'} \psi^L_{\b'}, 
\eear 
which is basic, closed and not exact. The cohomology therefore reads
\begin{align}
& H^p_\mathpzc{EQ}(\mathfrak{osp}(3|2) / \mathfrak{so}(2)\oplus \mathfrak{so}(1,1)) = \left \{
\begin{array}{lll}
\mathbb{R} & \quad & p = 0 \\
\mathbb{R}^2 & \quad & p \mbox{ even} \\
0 & \quad & p \mbox{ odd}.
\end{array}
\right. 
\end{align}

\subsubsection{$\mathfrak{osp}(4|2)$}

\noindent Finally, let us take a brief look at an interesting coset of $\mathfrak{osp}(4|2)$, namely $\mathfrak{osp}(4|2) / \mathfrak{u}(2)$. In this case the problem can be studied by considering the spinor representation of $\mathfrak{so}(1,3) \sim \mathfrak{so}(4) \cong \mathfrak{su}(2) \times \mathfrak{su}(2)$. In this formulation we have
\begin{eqnarray}
	\label{osp42A} && \mathpzc{T}^{[IJ]} \to \begin{cases}
		T^{(AB)} = (\sigma_{[IJ]})^{(AB)} \mathpzc{T}^{[IJ]} \\
		\tilde{T}^{(\dot{A} \dot{B})} = (\sigma_{[IJ]})^{(\dot{A} \dot{B})} \mathpzc{T}^{[IJ]}
	\end{cases} \ , \ A,B=1,2 \ , \\
	\label{osp42B} && \psi^I_\alpha \ \to \ \psi_{\alpha A \dot{A}} = (\sigma_I)_{A \dot{A}} \psi^{I}_\alpha \ ,
\end{eqnarray}
where $(\sigma^I)_{AB}$ and $(\sigma^I)_{\dot{A}\dot{B}}$ are the two copies of the Pauli matrices of $\mathfrak{su}(2) \times \mathfrak{su}(2)$, $\displaystyle (\sigma_{[IJ]})^{(AB)} = \left[ (\sigma_{I})^{A\dot{A}} , (\sigma_{J})_{\dot{A}}^{\ B} \right]$ and $\displaystyle (\sigma_{[IJ]})^{(\dot{A} \dot{B})} = \left[ (\sigma_{I})^{A\dot{A}} , (\sigma_{J})_ {A}^{\ \dot{B}} \right]$. The MC \eqref{ffB} then become
\begin{align}
\label{osp42C}
& d \mathpzc{V}_{(\a\b)} = \left( \psi \cdot \psi \right)_{(\alpha \beta)} + (\mathpzc{V} \wedge \mathpzc{V})_{(\a\b)}\,, \nonumber \\
& d \mathpzc{T}^{(AB)} = - \left( \psi \cdot \psi \right)^{(AB)} + (\mathpzc{T}\wedge \mathpzc{T})^{(AB)}\,, \nonumber \\
& d \tilde{\mathpzc{T}}^{(\dot{A}\dot{B})} = - \left( \psi \cdot \psi \right)^{(\dot{A}\dot{B})} + (\tilde{\mathpzc{T}}\wedge \tilde{\mathpzc{T}})^{(\dot{A}\dot{B})}\,, \\
& d \psi_{\a A \dot{A}} = \mathpzc{V}_{\a\b} \Omega^{\b\gamma} \psi_{\gamma A \dot{A}} + \sigma_{IA \dot{A}} \left( (\sigma^{[IJ]})_{(CD)}\mathpzc{T}^{(CD)} + (\sigma^{[IJ]})_{(\dot{C}\dot{D})}\mathpzc{T}^{(\dot{C}\dot{D})} \right) \eta_{JK} \sigma^{K A \dot{A}} \psi_{\a A \dot{A}} \nonumber \ .
\end{align}
Let us consider the coset $\displaystyle \mathfrak{osp}(4|2)/\mathfrak{su}(2)$: we can quotient out one of the two $\mathfrak{su}(2)$, for example the one generated by $\tilde{\mathpzc{T}}$. We immediately see that the bilinears $\left( \psi \cdot \psi \right)^{(\dot{A}\dot{B})} = - \nabla \tilde{\mathpzc{T}}^{(\dot{A}\dot{B})}$ become exact, with respect to non basic objects, hence are cohomology representatives of the coset algebra. The same holds for any power and product of these bilinears. Moreover, we have another cohomology representative which is given by \eqref{3form} but with just the non-modded out set of $\mathpzc{T}$'s:
\begin{equation}\label{osp42D}
	\omega^{(3)} = \left( \psi \cdot \psi \right)_{(\alpha \beta)} \mathpzc{V}^{(\a\b)} + \left( \psi \cdot \psi \right)^{(AB)} \mathpzc{T}_{(AB)} + 
(\mathpzc{V}\wedge \mathpzc{V}\wedge \mathpzc{V})_{\a\b} \Omega^{\a\b} + (\mathpzc{T}\wedge \mathpzc{T}\wedge \mathpzc{T})_{AB} \eta^{AB} \ .
\end{equation}
Hence, the cohomology is generated by $\displaystyle \left\lbrace 1 , \omega^{(3)} \right\rbrace \otimes \left\lbrace \left[ \left( \psi \cdot \psi \right)^{(\dot{A}\dot{B})} \right]^n \right\rbrace , \forall n \in \mathbb{N}$. The computation of the dimensions of the cohomology spaces is not difficult, but tedious since it heavily relies on the Fierz identities, hence it is not written here. Notice that the \virgolette finite part" of the cohomology (to be precise, its bosonic restriction) is exactly what is left from the bosonic quotient $\mathfrak{so}(4)/\mathfrak{su}(2) \cong \mathfrak{su}(2)$. Keeping in mind this observation, we will comment further on this in the next paragraph.

\noindent We could now proceed further by quotienting by another $\mathfrak{u}(1)$, in order to study the coset space $\displaystyle \mathfrak{osp}(4|2)/\mathfrak{u}(2)$. Given the previous results, the calculation is straightforward: in modding out with respect to $\mathfrak{u}(1)$, we can either embed it into the remaining $\mathfrak{su}(2)$, which is generated by the $\mathpzc{T}^{(AB)}$'s, or into $\mathfrak{sp}(2)$, which is generated by the $\mathpzc{V}^{(\a\b)}$'s. However, the two embeddings are equivalent since $\mathfrak{sp}(2) \cong \mathfrak{su}(2)$ at the level of complex algebras. Suppose then that we perform the quotient in the $\mathfrak{sp}(2)$ part, hence we immediately see that \eqref{osp42D} is no longer basic, hence it does not contribute to the cohomology. However, a 2-form as the one in \eqref{osp(2|2)V0} appears, making contribution to the cohomology. It follows that the cohomology of the coset $\displaystyle \mathfrak{osp}(4|2)/\mathfrak{u}(2)$ is generated by $\displaystyle \left\lbrace 1 , \mathpzc{R}^{(2)} \right\rbrace \otimes \left\lbrace \left[ \left( \psi \cdot \psi \right)^{(\dot{A}\dot{B})} \right]^n \right\rbrace , \forall n \in \mathbb{N}$. \\
A comment is now mandatory: as we already noticed at the end of the previous paragraph, the \virgolette finite part" of the cohomology (again, its bosonic restriction) corresponds to what is left from the bosonic quotient $\mathfrak{so}(4)/\mathfrak{u}(2) \cong \mathfrak{su}(2)/\mathfrak{u}(1)$. We can interpret this result, and the previous one, as follows: Fuks' theorem states that $H^p_{CE, \mathpzc{dif}} (\mathfrak{osp}(4|2)) = H^p_{CE, \mathpzc{dif}} (\mathfrak{so}(4)) $. When modding out the sub-algebra, we have found that bosonic restriction of the finite part is actually given by the coset of the (purely bosonic) subalgebra which is selected by Fuks' theorem. Notice that, not completely surprisingly, this holds true as long as we are embedding the sub-algebra in the part which actually contributes to the full cohomology of the algebra at the numerator. Indeed, we have seen in the $\mathfrak{u}(1|1)/\mathfrak{u}(1)$ example that if we embed the divisor subalgebra in the subalgebra not contributing to the cohomology, we obtain a different finite part. Hence, under the discussed assumption, it can be conjectured that, for example, if we consider the superalgebra $\mathfrak{osp}(n|m)$, given a subalgebra $\mathfrak{h}$, one will find that 
\begin{equation}\label{osp42E}
	\left[ H^p_{CE} \left( \frac{\mathfrak{osp}(n|m)}{\mathfrak{h}} \right) \right]_{FP} \cong \begin{cases}
		H^p_{CE} \left( \frac{\mathfrak{so}(n)}{\mathfrak{h}} \right) \ , \ \text{if} \ n \geq 2m \\
		H^p_{CE} \left( \frac{\mathfrak{sp}(m)}{\mathfrak{h}} \right) \ , \ \text{if} \ n < 2m
	\end{cases} \ ,
\end{equation}
where the subscript \virgolette FP" denotes the finite part of the cohomology. An evidence supporting this claim 
is provided by the Poincar\'e polynomials, which can be computed combining Fuks' results with 
\cite{greub}, in the case of equal rank pairs as follows 
\begin{eqnarray}
\label{polliA}
 \mathpzc{P}_{{\mathfrak{osp}(n|m)}/{\mathfrak{h}}} (t) = \frac{\mathpzc{P}'_{{\mathfrak{osp}(n|m)}} (t) }{
 \mathpzc{P}'_{{\mathfrak{h}}}(t)}
\end{eqnarray}
where the prime denotes the augmented power by one for all powers in the Poincar\'e series. 
It would be interesting to verify if this holds for cosets of the other basic Lie superalgebras as well as to improve general results comprising also quotient spaces, as in \cite{greub}.

\subsection{Cosets of $\mathfrak{osp}(1|4)$: $D=4$, $\mathcal{N}=1$ Anti de Sitter Superspace}

\noindent It is well-known that the ordinary anti de Sitter spacetimes $AdS_D$ in $D$-dimensions can be obtained starting from the Lie groups $SO(2, D-1)$ and $SO (1, D-1)$ as the coset manifold $SO(2, D-1) / SO(1,D-1)$, for example the $AdS_4$ is obtained by taking the quotient of the Lie group $SO(2,3)$ by the Lorentz group $SO(1,3)$ (see \cite{cube} for a complete discussion on the present case in relation to supergravity and in particular in relation with Chevalley-Eilenberg cohomology. In \cite{cube} the computation of the easiest CE cohomology has been performed) . This construction can be generalized to a coset superspace as to obtain the superspace extension of the anti-de Sitter spacetimes. Namely, in this section we are interested into computing the equivariant cohomology of one such construction, namely the $D=4$, $\mathcal{N}=1$ anti-de Sitter superspace $AdS_{4|4}$ realized as the quotient supermanifold $OSp(1|4)/SO(1,3)$. At the level of the Lie superalegbras one starts analyzing the $\mathfrak{osp}(1|4),$ of dimension $10|4$, whose reduced Lie algebra is $\mathfrak{osp}(1|4)_0 = \mathfrak{sp}(4, \mathbb{R})$. Using that $\mathfrak{sp} (4, \mathbb{R}) \cong \mathfrak{so}(2,3)$ one can trace back the quotient yielding the anti de Sitter $4$-space $AdS_4$ at the level of the groups. Notice that the quotient manifold $OSp(1|4) / SO(1,3)$ has dimension $4|4$, therefore it is $\mathcal{N}=1$ (minimal) supersymmetric extension for the $AdS_4$ and we call it $AdS_{4|4}$. We will denote the corresponding coset at the Lie superalgebra level $\mathfrak{ads}_{4|4} \defeq \mathfrak{osp}(1|4) / \mathfrak{so}(1,3)$.\\ 

Let us start analyzing the the Chevalley-Eilenberg cohomology of $\mathfrak{osp}(1|4)$. At the level of the Poincar\'e polynomial we have
\bear
\mathpzc{P}_{\mathfrak{osp}(1|4)} [t]= \mathpzc{P}_{\mathfrak{sp}(4, \mathbb{R})} [t] = 1 - t^3 - t^7 + t^{10}.
\eear
Introducing a set of gamma matrices $\gamma^a_{\alpha \beta}$ for $a = 0, \ldots, 9$ and $\alpha, \beta = 1, \ldots, 4$ we represent the Maurer-Cartan odd forms by bi-spinors as follows
\bear
\mathpzc{V}^a = \gamma^a_{\alpha \beta} \mathpzc{V}^{\alpha \beta} \ , \ a=1,\ldots,10, \quad \alpha, \beta=1,\ldots,4. 
\eear
Notice that is consistent as long as the indices $\a, \b$ are symmetrized, \emph{i.e.} $\mathpzc{V}^{\a \b} = \mathpzc{V}^{(\a \b)}$, as to yield $10$ components. Further, we use the (standard) symplectic matrix $\Omega_{\a \b}$ and its inverse $\Omega^{\a \b} $ to lower and raise indices. This representation is particularly convenient, as the Maurer-Cartan equations read
\begin{align}
	\nonumber & d \mathpzc{V}_{\alpha \beta} = \psi_\alpha \psi_\beta + \left( \mathpzc{V} \Omega \mathpzc{V} \right)_{\alpha \beta}, \\
	\label{OSPN4F} & d \psi_\alpha  = \left( \mathpzc{V} \Omega \psi \right)_\alpha,
\end{align}
having introduced the (even) vielbeins $\psi^\alpha$ as well and where we have made use of the notation $\displaystyle \left( \mathpzc{V} \Omega \mathpzc{V} \right)_{\alpha \beta} = \mathpzc{V}_{\alpha \gamma} \Omega^{\gamma \delta} \mathpzc{V}_{\delta \beta} $ and $\displaystyle \left( \mathpzc{V} \Omega \psi \right)_\alpha = \mathpzc{V}_{\alpha \beta} \Omega^{\beta \gamma} \psi_\gamma.$ Let us look for the 3-form explicitely: the most general 3-form reads
\bear
\omega^{(3|0)} = c_1 \left( \mathpzc{V}^{\alpha_1 \beta_1} \Omega_{\beta_1 \alpha_2} \mathpzc{V}^{\alpha_2 \beta_2} \Omega_{\beta_2 \alpha_3} \mathpzc{V}^{\alpha_3 \beta_3} \Omega_{\beta_3 \alpha_1} \right) + c_2 \mathpzc{V}^{\alpha \beta} \psi_\alpha \psi_\beta,
\eear
where $c_1$ and $c_2$ are constants coefficient. We shorten the previous expression by $\omega^{(3)} \defeq c_1 \mathpzc{V}^{(3)} + c_2 \mathpzc{V} \psi^{(2)}.$ By compatibility with the cohomology of the reduced algebra $\mathfrak{sp}(4, \mathbb{R})$ we conclude that $c_1 \neq 0$, and in particular, we put $c_1 = 1$. Imposing the closure condition $d\omega^{(3)} = 0$ we fix the second coefficient: 
\begin{equation}\label{OSPN4H}
	0 = d \omega^{(3)} = 3 \left[ \left( \psi^{\alpha_1} \psi^{\beta_1} + \left( \mathpzc{V} \Omega \mathpzc{V} \right)^{\alpha_1 \beta_1} \right) \Omega_{\beta_1 \alpha_2} V^{\alpha_2 \beta_2} \Omega_{\beta_2 \alpha_3} \mathpzc{V}^{\alpha_3 \beta_3} \Omega_{\beta_3 \alpha_1} \right] + $$ $$ + c_2 \left[ \psi^{\alpha} \psi^\beta \psi_\alpha \psi_\beta - 2 \mathpzc{V}^{\alpha \beta} \mathpzc{V}_{\alpha \gamma} \Omega^{\gamma \delta} \psi_\delta \psi_\beta \right].
\end{equation}
Let us look at the terms in this expression: the second term, namely the one proportional to $\mathpzc{V}^4$ is zero by trace identity, indeed we can write $\mathpzc{V}^3 \mathpzc{V} = - \mathpzc{V} \mathpzc{V}^3$, but on the other hand, by cyclicity we have $\mathpzc{V}^3 \mathpzc{V} = \mathpzc{V}\mathpzc{V}^3$. The third term, namely the one proportional to $\psi^4$, is zero since $\psi^\alpha \psi_\alpha = \psi^\alpha \Omega_{\alpha \beta} \psi^\beta = 0$, being the $\psi$'s even and $\Omega$ antisymmetric. This allows us to fix $c_2 = 3/2$ as to get that the first cancel the last term and obtaining a closed form. Further, in order to show that $\omega^{(3)}$ is not exact, we have to consider the most general even $2$-form and show that its Chevalley-Eilenberg differential cannot generate $\omega^{(3)}$. However a crucial observation simplifies the job: we cannot construct a (non-zero) $2$-form which is a \emph{singlet}, \emph{i.e.}\ having all of the indices contracted (the only case would be $V^{\alpha \beta} V_{\alpha \beta} + \psi^\alpha \psi_\alpha$ which is equal to zero, as shown above). Hence we have (after multiplying by an overall factor)
\begin{equation}\label{OSPN4I}
	H^{3}_{CE} \left( \mathfrak{osp} \left( 1|4 \right) \right) = \left\lbrace \frac{1}{3} \left( V^{\alpha_1 \beta_1} \Omega_{\beta_1 \alpha_2} V^{\alpha_2 \beta_2} \Omega_{\beta_2 \alpha_3} V^{\alpha_3 \beta_3} \Omega_{\beta_3 \alpha_1} \right) + \frac{1}{2} V^{\alpha \beta} \psi_\alpha \psi_\beta \right\rbrace \ .
\end{equation}
With completely analogous arguments we can construct the most general odd $7$-form as
\begin{eqnarray}
	\nonumber \omega^{(7)} &=& c_1 \left( \mathpzc{V}^{\alpha_1 \beta_1} \Omega_{\beta_1 \alpha_2} \mathpzc{V}^{\alpha_2 \beta_2} \Omega_{\beta_2 \alpha_3} \mathpzc{V}^{\alpha_3 \beta_3} \Omega_{\beta_3 \alpha_4} V^{\alpha_4 \beta_4} \Omega_{\beta_4 \alpha_5} \mathpzc{V}^{\alpha_5 \beta_5} \Omega_{\beta_5 \alpha_6} \mathpzc{V}^{\alpha_6 \beta_6} \Omega_{\beta_6 \alpha_7} V^{\alpha_7 \beta_7} \Omega_{\beta_7 \alpha_1} \right) + \\
	\nonumber &+& c_2 \left( V^{\alpha_1 \beta_1} \Omega_{\beta_1 \alpha_2} \mathpzc{V}^{\alpha_2 \beta_2} \Omega_{\beta_2 \alpha_3} \mathpzc{V}^{\alpha_3 \beta_3} \Omega_{\beta_3 \alpha_4} \mathpzc{V}^{\alpha_4 \beta_4} \Omega_{\beta_4 \alpha_5} \mathpzc{V}^{\alpha_5 \beta_5} \right) \psi_{\alpha_1} \psi_{\alpha_5} + \\
	\nonumber &+& c_3 \left( \mathpzc{V}s^{\alpha_1 \beta_1} \Omega_{\beta_1 \alpha_2} \mathpzc{V}^{\alpha_2 \beta_2} \Omega_{\beta_2 \alpha_3} \mathpzc{V}^{\alpha_3 \beta_3} \Omega_{\beta_3 \alpha_1} \right) \left( \mathpzc{V}^{\alpha_1 \mu} \Omega_{\mu \nu} \mathpzc{V}^{\nu \alpha_2} \right) \psi_{\alpha_1} \psi_{\alpha_2} + \\
	\label{OSPN4J} &+& c_4 \left( \mathpzc{V}^{\alpha_1 \beta} \Omega_{\beta \gamma} \mathpzc{V}^{\gamma \alpha_2} \right) \mathpzc{V}^{\alpha_3 \alpha_4} \psi_{\alpha_1} \psi_{\alpha_2} \psi_{\alpha_3} \psi_{\alpha_4}.
\end{eqnarray}
We note that we do not have a term of the form $\mathpzc{V} \psi^6$ since it would be trivially 0, as can be checked. We can write $\omega^{(7)}$ in a more compact way as
\begin{equation}\label{OSPN4K}
	\omega^{(7)} = c_1 \mathpzc{V}^7 + c_ 2\left( \mathpzc{V}^5 \right)^{\alpha \beta} \psi_\alpha \psi_\beta + c_3 \mathpzc{V}^3 \left( \mathpzc{V}^2 \right)^{\alpha \beta} \psi_\alpha \psi_\beta + c_4 \left( \mathpzc{V}^2 \right)^{\alpha \beta} \mathpzc{V}^{\gamma \delta} \psi_\alpha \psi_\beta \psi_\gamma \psi_\delta \ ,
\end{equation}
where the contractions are omitted. Again by compatibility with the reduced algebra cohomology, we need to have $c_1\neq 0$. The remaining coefficients $c_2, c_3, c_4$ can be easily fixed imposing $d \omega^{(7)} = 0$: again, as above, the resulting form will not be exact since it is not possible to have a non-trivial singlet represented by an even $6$-form. \\
Finally the top representative in the cohomology, the form $\omega^{(10)}$ is simply given given by the multiplication
\begin{equation}\label{OSPN4L}
	\omega^{(10)} = \omega^{(3)} \wedge \omega^{(7)},
\end{equation}
exploiting the ring structure of the cohomology. Notice that $\omega^{(10)}$ is non-zero since, for example, the term of the form $\mathpzc{V}^3 \wedge \mathpzc{V}^7$ is non-vanishing, and since either $\omega^{(3)}$ or $\omega^{(7)}$ are closed and non-exact it follows that $\omega^{(10)}$ is closed and non-exact as well.\\

We now pass to study the equivariant cohomology of the coset superspace $\mathfrak{ads}_{4|4} = \mathfrak{osp}(1|4) / \mathfrak{so}(1,3)$. In order to do so, we have to \virgolette split'' the Maurer-Cartan forms $\mathpzc{V}^{\a \b}$ coming from the $\mathfrak{sp}(4, \mathbb{R}) \subset \mathfrak{osp}(1|4)$ into the coset Maurer-Cartan forms (vielbeins) and those coming from $\mathfrak{so}(1,3)$ (connections). Again, making use of the gamma matrices, \emph{i.e.}\ of the spin structure, we can decompose the vielbeins as
\bear \label{decads}
	\mathpzc{V}_{(\alpha \beta)} = \gamma^a_{(\alpha \beta)} \mathpzc{V}_a + \gamma^{[a b]}_{(\alpha \beta)} \mathpzc{V}_{[ab]}, 
\eear 
for $a = 1,\ldots,4$  and $\alpha=1,\ldots,4$, where the $\mathpzc{V}^a$ are the four vierbein of the coset space that lifts to $AdS_4$ and $\mathpzc{V}_{[ab]}$ are the six vielbeins of $\mathfrak{so}(1,3)$. The Poincar\'e polynomial can be computed using the result by \cite{greub} - notice that both the algebras involved have the same rank, actually 2 - and it reads
\bear
\mathpzc{P}_{\mathfrak{ads}_{4|4}} [t] = \frac{\left( 1 - t^4 \right) \left( 1 - t^8 \right)}{\left( 1 - t^4 \right)^2} = 1 + t^4.
\eear
We therefore expect a single equivariant cohomology class at degree 4, besides the constants. In particular, we expect this to be related to the \virgolette volume form'' $\omega^{(4)}_{\mathfrak{ads}_4} $ coming from the $AdS_4$ space. Using the above decomposition \eqref{decads} and the previously obtained Maurer-Cartan equations \eqref{OSPN4F} one gets the following Maurer-Cartan equations
\bear
&& \mathcal{D} \mathpzc{V}_a = \psi \gamma_a \psi\,, ~~~~\nonumber \\
&& \mathcal{D} \mathpzc{\psi}_\alpha = \mathpzc{V}_a \gamma^a \psi \,, ~~~~\nonumber \\
&& R_{ab} \equiv d  \mathpzc{V}_{[ab]} + ( \mathpzc{V} \wedge \mathpzc{V})_{[ab]} = \psi \gamma_{[ab]} \psi
\eear
where the covariant derivative $\mathcal{D}$ is with respect to the connection $ \mathpzc{V}_{[ab]}$ of the 
subgroup $\mathfrak{so}(1,3)$. 

Working as above, we have that the most general even $4$-singlet reads
\bear
	\omega^{(4|0)} = c_1 \epsilon_{abcd} \mathpzc{V}^a \mathpzc{V}^b \mathpzc{V}^c \mathpzc{V}^d + c_2 \mathpzc{V}^a \mathpzc{V}^b \left( \psi \gamma_{ab} \psi \right).
\eear
Notice that there cannot be terms of the form $ \psi^4 = \left( \psi \gamma^{ab} \psi \right) \left( \psi \gamma_{ab} \psi \right) $, since they vanish because of the Fierz identities. As above, we have that $c_1\neq0$ by compatibility with the cohomology of the reduced algebra $\omega^{(4)}_{\mathfrak{ads}_4} = \epsilon_{abcd} V^a V^b V^c V^d$. Hence we can fix $c_1=1$ without loss of generality. The coefficient $c_2$ is fixed by imposing that $\mathcal{D} \omega^{(4)} = 0$:
\begin{equation}
	0 = \mathcal{D} \omega^{(4)} = 4 \epsilon_{abcd} \left( \psi \gamma^a \psi \right) \mathpzc{V}^b \mathpzc{V}^c \mathpzc{V}^d + 2 c_2 \left[ \psi \gamma^a \psi \mathpzc{V}^b \left( \psi \gamma_{ab} \psi \right) + \mathpzc{V}^a \mathpzc{V}^b \left( \left( \mathpzc{V}^c \gamma_c \psi \right) \gamma_{ab} \psi \right) \right] \ .
\end{equation}
The second term in the sum vanishes because of Fierz identities, while the last term, after using $\gamma$ matrices properties, cancels the first one upon fixing $\displaystyle c_2=-2$. Finally, we can conclude that $\omega^{(4)}$ is not exact, since, once again, it is not possible to write an odd $3$-singlet that generates the term $\mathpzc{V}^4$. Hence we have
\begin{equation}
	H^{4}_{\mathpzc{EQ}} \left(\mathfrak{ads}_{4|4} \right) = \mathbb{R}\cdot \left\lbrace \epsilon_{abcd} \mathpzc{V}^a \mathpzc{V}^b \mathpzc{V}^c \mathpzc{V}^d -2 \mathpzc{V}^a \mathpzc{V}^b \left( \psi \gamma_{ab} \psi \right) \right\rbrace .
\end{equation}
All in all we have:
\bear
H^{p}_{\mathpzc{EQ}} \left (\mathfrak{ads}_{4|4} \right ) = \left \{
\begin{array}{lll}
\mathbb{R} & \quad & p = 0, 4 \\
0 & \quad &  \mbox{else}.
\end{array}
\right.
\eear
We conclude with the integral form Chevalley-Eilenberg cohomology. 
As discussed in the previous section, by the isomorphism, we have two 
cohomology classes at picture four, the maximal picture degree. They have the explicit 
expressions 
\begin{eqnarray}
\label{cippolippoA}
H^{(0|4)} \left (\mathfrak{ads}_{4|4} \right ) &=& \mathbb{R} \cdot 
 \left\lbrace 2 \mathpzc{V}^a \mathpzc{V}^b \iota_{\pi Q} \gamma_{ab} \iota_{\pi Q} \delta^4(\psi) + \delta^4(\psi)\right\rbrace \nonumber 
\\
H^{(4|4)} \left (\mathfrak{ads}_{4|4} \right ) &=& \mathbb{R} \cdot 
 \left\lbrace \epsilon_{abcd} \mathpzc{V}^a \mathpzc{V}^b \mathpzc{V}^c \mathpzc{V}^d  \delta^4(\psi) 
 \right\rbrace
\end{eqnarray}
matching again the cohomology for superforms.

\section{Conclusions and Outlook}

\noindent The present work spawns from the observation that since Lie superalgebra cohomology is nothing but a straightforward generalization of ordinary Lie algebra cohomology, it is not capable to account for objects different than differential forms on the corresponding Lie supergroup. On the other hard, it is well-know that in order to make a meaningful connection with integration theory, when working on supermanifolds, differential forms has to be supplemented by integral forms, whose geometry is not at all captured by Chevalley-Eilenberg cohomology. \\
To this end, after reviewing Chevalley-Eilenberg cohomology for ordinary Lie algebras and Lie superalgebras and its relations to forms on the corresponding Lie groups or Lie supergroups, we extend the notion of Chevalley-Eilenberg cochains as to include also \emph{integral forms} and we define a corresponding cohomology. We thus show a duality between the ordinary Chevalley-Eilenberg cohomology for a certain Lie superalgebra - which looks at forms on the corresponding Lie supergroup - and this newly defined (Chevalley-Eilenberg) cohomology accounting for integral forms instead. We observe that, most notably - and differently from de Rham cohomology -, this cohomology always feature the true analog of a top-form, a Berezinian form appearing in the integral form complex.	\\
Nonetheless, beside general results, a great deal of focus in this paper is on explicit direct computations: in particular, we provide explicit expressions for cocycles of Lie superalgebras of physical interest, namely supertranslations of flat superspaces and classical Lie superalgebras, up to dimension 4, in terms of their Maurer-Cartan forms.  \\
The second part of the paper is devoted to equivariant Chevalley-Eilenberg cohomology, which is related to the (super)symmetries of coset supermanifolds, which provides very important backgrounds for supergravity and superstring theories. Again, several example up to dimension 4 are studied and explicit expressions for their cocycles are provided, culminating with the case of super anti-de Sitter space $AdS_{4|4}$. Here, a mixture of techniques have been exploited, spanning from Poincar\'{e} polynomials computations for equal rank pairs to brute force computations.\\
We remark that our analysis have uncovered new cocycles spawning from fermionic generators - both in ordinary and equivariant Chevalley-Eilenberg cohomology - and several characteristic examples of infinite dimensional cohomology. In hoping that the present results might come useful to understand the geometry of supergravity and string backgrounds and the mathematics behind it, we stress though, that this research scenario looking at relating Chevalley-Eilenberg cohomology and the extended geometry of forms on supermanifolds is far from being exhausted. Indeed, just as an example, in the present paper we have only hinted at pseudoforms, which nonetheless plays an important role both in the integration theory on superspaces and in its applications: it is legit to ask it they can be fitted in the picture we have presented and which role they play. We will address this problem in a forthcoming paper \cite{CCGN2}, arguing that pseudoforms are indeed crucial to understand the general structure of the cohomology.

\section*{Acknowledgements}
\noindent This work has been partially supported by Universit\`a del Piemonte Orientale research funds. We thank L. Castellani and P. Aschieri for many useful discussions.

\appendix
\section{The Unitary Lie Superalgebra $\mathfrak{u} (n|m)$}
\label{unit}

\noindent Following \cite{CouZha}, in order to introduce the Lie superalgebra $\mathfrak{u} (n|m)$ one can start with the ordinary Hermitian vector space $(\mathbb{C}^{n+m}, \langle \cdot , \cdot \rangle_{\mathbb{C}^{n+m}})$, where $\langle \cdot , \cdot \rangle_{\mathbb{C}^{n|m}} $ is the standard Hermitian product: promoting $\mathbb{C}^{n+m}$ to a vector superspace $\mathbb{C}^{n|m}$ using the $\mathbb{Z}_2$-gradation, one then defines the super Hermitian forms on $\mathbb{C}^{n|m}$ as $\langle u , v \rangle_{\mathbb{C}^{n|m}} \defeq (-1)^{|u| |v|} \langle u, v\rangle_{\mathbb{C}^{n+m}},$ where $u$ and $v$ are $\mathbb{Z}_2$-homogeneous vectors in $\mathbb{C}^{n|m}$. Notice that $\langle u, v\rangle_{\mathbb{C}^{n|m}} = (-1)^{|u| |v|} \overline{\langle v, u \rangle }_{\mathbb{C}^{n|m}}$, so that this bilinear form is indeed Hermitian in the usual sense. Using this, the \emph{superadjoint} of an endomorphism $A \in End (\mathbb{C}^{n|m})$ is naturally defined as $ \langle A u ,  v \rangle_{\mathbb{C}^{n|m}} = (-1)^{|A||u|}\langle u, A^\ast v \rangle_{\mathbb{C}^{n|m}}$ and it is easy to see that $T^\ast = i^{|T|} T^\dagger$, where the map $T \mapsto T^\dagger$ does not involve the supertransposition, but just the ordinary transposition, \emph{i.e.} $T^\dagger$ is the usual adjoint with respect to standard Hermitian form on $\mathbb{C}^{n+m}$. These definitions leads immediately to $[A, B]^\ast = - [A^\ast, B^\ast]$, which spell out the relations between superadjoint and the supercommutator, which is what is needed in order to define a unitary representation of a Lie superalgebra: in particular if $\rho : \mathfrak{g} \rightarrow End (\mathbb{C}^{n|m})$ is a representation of $\mathfrak{g}$, we will say that it is a \emph{unitary} representation if $\rho (A)^\ast = - \rho (A)$ for $A \in \mathfrak{g}.$ For the case of supermatrices $X \in \mathfrak{gl} (n|m, \mathbb{C})$, this leads to the definition   
\bear
\mathfrak{u} (n|m) \defeq \left \{ X \in \mathfrak{gl} (n|m, \mathbb{C}): X^\ast = -X \right \}.
\eear
Realizing the above conditions in terms of matrices $\mathbb{C}^{n|m} \times \mathbb{C}^{n|m}$, one finds
\bear
X = \left ( \begin{array}{c|c}
A & \Theta \\
\hline 
- i \Theta^\dagger & B
\end{array}
\right ),
\eear
for $A \in \mathfrak{u}(p)$ and $B \in \mathfrak{u} (q)$, so that $A^\dagger = -A$ and $B^\dagger = -B$, and $\Theta \in Hom (\Pi \mathbb{C}^{m} , \mathbb{C}^{n}),$ \emph{i.e.} an odd matrix. This easily yield 
\bear 
\dim_{\mathbb{R}} \mathfrak{u} (n|m) = n^2 + m^2 | 2nm.
\eear

\section{The Orthosymplectic Lie Superalgebra $\mathfrak{osp}(n|2m)$} \label{appB}

\noindent Working in the most general setting following again \cite{CouZha}, given a number field $k$ of characteristic 0 the natural representation of the general linear Lie superalgebra $\mathfrak{gl} (n|m, k)$ on the vector superspace $k^{n|m}$ can be extended to a representation acting on the tensor algebra $\mathpzc{T}ens (k^{n|m}) \defeq \bigoplus_{n\geq 0} (k^{n|m})^{\otimes n}$, upon using the graded Leibniz rule, \emph{i.e.}\ for $A \in \mathfrak{gl}(n|m, k)$ and $v_i \in k^{n|m}$ homogeneous vectors one has
\begin{align}
A \cdot (v_1 \otimes \ldots \otimes v_n \otimes \ldots ) = &(A \cdot v_1) \otimes v_2 \otimes \ldots \otimes v_n \otimes \ldots + \nonumber \\
& + (-1)^{|A||v_1|} v_1 \otimes (A\cdot v_2) \otimes v_3 \otimes \ldots \otimes v_n \otimes \ldots + \ldots + \nonumber \\
& + (-1)^{\sum_{i=1}^{n-1}  |A||v_i|} v_1 \otimes \ldots \otimes v_{n-1} \otimes (A\cdot v_n) \otimes v_{n+1} \otimes \ldots.  
\end{align}
Choosing $k=\mathbb{R}$, it is possible to introduce a bilinear form on $G: \mathbb{R}^{n|2m} \otimes \mathbb{R}^{n|2m} \rightarrow \mathbb{R}$, such that for the standard basis $\mbox{Span}_{\mathbb{R}} \{e_i \} = \mathbb{R}^{n|2m}$, one has 
\bear
G (e_i \otimes e_j ) = {g}_{ij} \quad \mbox{with} \quad {g}_{ij} = \left ( \begin{array}{c|cc}
{1}_{n\times n} & {0}_{m \times n} & {0}_{m \times n} \\
\hline
{0}_{m \times n} & {0}_{m \times m} & {1}_{m \times m} \\
{0}_{m\times n} & -{1}_{m \times m} & {0}_{m \times m} 
\end{array}
\right ).
\eear 
For short, we define
\bear
G \defeq \left ( 
\begin{array}{c|c}
1_n & \\
\hline
   & J_{2m}
\end{array}
\right ) \quad \mbox {with} \quad J_{2m} \defeq \left ( \begin{array}{cc}
0_m & 1_m \\
-1_m & 0_m 
\end{array}
\right ).
\eear
Here $J$ is just the \emph{standard symplectic matrix}, which has the property that $J = - J^t.$ \\
The orthosymplectic Lie superalgebra can therefore be defined as 
\bear
\mathfrak{osp}(n|2m) \defeq \{ X \in \mathfrak{gl} (n|2m, \mathbb{R}) : G (X \cdot ( v_1 \otimes v_2)) = 0, \; \forall v_1, v_2 \in \mathbb{R}^{n|2m} \}.
\eear
One sees that, unraveling the above definition, one gets the following condition on $X \in \mathfrak{gl}(n|2m, \mathbb{R}):$
\bear
X^tG + G X = 0.
\eear
In turn, writing $X$ in block-form
\bear
X \defeq \left ( \begin{array}{c|c}
A & \Phi \\
\hline
\Psi & B
\end{array}
\right ) 
\eear
for $A \in Hom_{\mathbb{R}} (\mathbb{R}^{n|0}, \mathbb{R}^{n|0})$, $B \in Hom_\mathbb{R} (\mathbb{R}^{0|2m}, \mathbb{R}^{0|2m})$ even matrices and $\Phi \in Hom_{\mathbb{R}} (\mathbb{R}^{0|2m} , \mathbb{R}^{n|0})$ and $\Psi \in Hom_{\mathbb{R}} (\mathbb{R}^{n|0} , \mathbb{R}^{0|2m})$ odd matrices one finds the conditions
\begin{align} \label{relosp}
A^{t} = - A , \qquad B^t =  JBJ, \qquad \Psi = J \Phi^t,
\end{align}
having used that $-J^t = J$ in the relation for $B$, so that the generic element of the superalgebra can be written as  
\bear
\mathfrak{osp} (n|2m) \owns X \defeq \left ( \begin{array}{c|c}
A & \Phi \\
\hline
J \Phi^t & B
\end{array}
\right ), 
\eear
with $A \in \mathfrak{so} (n, \mathbb{R})$ and $B \in \mathfrak{sp} (2n, \mathbb{R})$, which explains the denomination \emph{orthosymplectic}. Also, the above conditions makes it easy to count the dimensions of this Lie superalgebra, namely one finds
\bear
\dim_\mathbb{R} \mathfrak{osp} (n|2m) = \frac{1}{2} n(n-1) + m (2m + 1) | 2mn.
\eear

\vfill
\eject

\end{document}